\documentstyle[12pt,fullpage,epsf]{article}
\let\sf=\rm
\def\d{\delta}
\def\e{\epsilon}
\def\g{\gamma}
\def\h{\eta}
\def\j{\psi}
\def\l{\lambda}
\def\m{\mu}
\def\t{\tau}
\def\del{\partial}
\def\cl{{\cal L}}
\def\half{\frac{1}{2}}

\def\eq{\begin{equation}}
\def\eqe{\end{equation}}
\def\eqa{\begin{eqnarray}}
\def\eqae{\end{eqnarray}}

\begin{document}
\begin{titlepage}
\renewcommand{\thefootnote}{\fnsymbol{footnote}}
\begin{flushleft} 
ITP-SB 98-9
\hfill hep-th/9802074 \\
TUW 98-01
\end{flushleft}  
\begin{center}  \vfil 
{\LARGE\bf
Topological  boundary conditions, the BPS\\ bound, and 
elimination of ambiguities in the\\[5pt]
quantum mass of solitons\footnote{The work is supported by
NSF grant PHY97-22101.} 
}
\bigskip\vfil
{\Large Horatiu Nastase, Misha Stephanov, Peter van Nieuwenhuizen}\\[5pt]
{Institute for Theoretical Physics},\\
{S.U.N.Y. at Stony Brook, NY 11794-3840, USA}\\[5pt]
{and}\\[5pt] 
{\Large Anton Rebhan}\\[5pt]
{Institut f\"{u}r Theoretische Physik, 
Technische Universit\"{a}t Wien,}\\
{Wiedner Hauptstr. 8-10, A-1040 Vienna, Austria}
\end{center}
\vfil
\centerline{\large\bf  Abstract}\vspace{.5cm}
We 
fix the long-standing ambiguity in the 1-loop contribution to the 
mass of a 1+1-dimen\-sional supersymmetric
soliton by adopting a set of boundary conditions which follow from the 
symmetries of the action and which depend only on the topology of the sector 
considered, and by
invoking a physical principle that ought to hold generally in
quantum field theories with a 
topological sector: for vanishing mass and other dimensionful constants,
the vacuum energies in the trivial and topological sectors have to
become equal.
In the two-dimensional $N=1$ supersymmetric case we find a result
which for the supersymmetric
sine-Gordon model agrees with the known exact solution of the
S-matrix but seems to violate the BPS bound. We analyze 
the nontrivial relation between the quantum soliton mass and the
quantum BPS bound and find a resolution.
For $N=2$ supersymmetric theories, there are no one-loop corrections
to the soliton mass and to the central charge (and also no ambiguities)
so that the BPS bound is always saturated. 
Beyond 1-loop there are no
ambiguities in any theory, which we explicitly check by a 2-loop calculation
in the sine-Gordon model.
\renewcommand{\thefootnote}{\arabic{footnote}}
\setcounter{footnote}{0}
\end{titlepage}

\section{Introduction}
The calculation of quantum corrections to the mass of solitonic
objects has been a subject of intense interest in the past \cite{raj} and has 
recently been
revived in the light of a recent breakthrough in our
understanding of non-perturbative dynamics in 3+1-dimensional
supersymmetric (susy) gauge
theories~\cite{SW}. One of the most important ingredients of the
non-perturbative analysis of such theories is the duality between
extended solitonic objects, such as monopoles or dyons, and
point-like particles. Another important ingredient is the concept of
the BPS spectrum --- the particles whose masses are
proportional to their charges.  Due to the supersymmetry algebra
BPS states may be annihilated by the action of some of the
supersymmetry generators and hence give rise to a smaller number
of superpartners (``multiplet shortening'').  
Therefore the BPS value of the mass becomes a
qualitative, rather than just a quantitative property \cite{WO}.

Susy models in 1+1 dimensions provide a valuable
nontrivial testing ground for these concepts. Solitons in these
models are examples of BPS states whose mass  at the classical level
is proportional to the topological charge. 
The calculation
of quantum corrections to the masses of these states has been
subject to a long controversy.
A number of one-loop calculations
have been performed yielding different contradicting
results \cite{dAdV,sch,Rouh,kar1,imb1,Y,CM,Uch,uchi1,uchi2,rvn}.

The one-loop corrections to the mass 
of a soliton are given by
\begin{equation}
M_{sol}^{(1)}={1\over2}\sum \left(\omega^B-\omega^F\right) - 
{1\over2}\sum\left(\tilde{\omega}^B-\tilde{\omega}^F\right)+ \d M
\label{mass}
\end{equation}
where $\omega^{B,F}$ are the energies of the small 
bosonic (fermionic) fluctuations about the 
classical soliton solution, $\tilde{\omega}$ are the corresponding
energies of the linearized
theory in the trivial vacuum,
while $\d M$ is the counter\-term which can be obtained
from the expression for the classical
mass of the soliton in terms of unrenormalized parameters by expanding
into renormalized ones \cite{col}. 
In order to make the above sums well defined,
spatial boundaries are temporarily introduced to make
the entire spectrum discrete.

One can identify two sources of ambiguities. As was discussed in
\cite{sch}, imposing different spatial boundary conditions on the
small quantum fluctuations around the classical soliton gives
different, sometimes even ultraviolet divergent, results. In this
paper we present an analysis which answers the question: which
boundary conditions are to be used in the one-loop calculation?  The
answer to this question can be found if one re-examines the original
formulation of the problem. We consider the vacuum energy as a
functional of the boundary conditions. We then single out a class of
boundary conditions which do not introduce surface effects --- the
topological boundary conditions. They close the system on
itself. There is a trivial (periodic) as well as a topologically
non-trivial (with a Moebius-like twist) way of doing this.
 This definition of boundary conditions does not rely on a
semiclassical loop expansion. We do not separate the classical part of
the field from its quantum fluctuations; rather, the boundary
conditions are imposed on the whole field. One then infers the correct
boundary conditions for the classical part as well as for the quantum
fluctuations from this single general condition.

Another source of ambiguity, as was pointed out recently \cite{rvn}, is the
choice of the ultraviolet regularization scheme. The dependence on
the choice of regularization scheme can be understood as a peculiar
property of those quantities, such as the soliton mass, which involve a
comparison between two sectors with different boundary conditions, i.e.,
different topology. Indeed, the difference of the vacuum
energies in the two sectors measures the mass of the soliton.
The one-loop correction is then given by a sum
over zero-point frequencies in the soliton sector
which is quadratically divergent. 
A similar sum in another (the trivial) sector is to be subtracted
in order to get an expression which is finite if written in terms
of the renormalized parameters. It turns out that due to the bad
ultraviolet behavior of both sums the result depends on the
choice of the cut-off \cite{rvn}. Cutting both sums off
at the same energy 
in both sectors \cite{kar1,imb1,Y,CM} or taking equal numbers of modes
in both sectors \cite{dhn1,sch,uchi1,uchi2} leads to different results.
To add to the confusion, some authors do not include bound states and/or
zero modes when they consider equal numbers of states in both 
sectors.\footnote{See for example the textbook \cite{raj},
eq.~(5.60). Actually in this
reference the result of \cite{dhn1} is obtained but this requires neglecting
a boundary term in the partial integration of (5.63).}
In this paper
we propose a simple way of reducing the ultraviolet divergence
of the sums over the zero-point energies, which eliminates
the sensitivity to the ultraviolet cutoff. Instead of calculating
the sums we calculate their
derivative with respect to the physical mass scale in the theory. 
The constant of integration can
be fixed by using the following observation: the vacuum energy
should not depend on the topology when the mass is zero.
This is the physical principle which allows us to perform the
calculation unambiguously. It should be viewed as a renormalization
condition. 

From a practical point of view we need to use our condition that the
vacuum energy functional does not depend on topology at zero mass only
in the one-loop calculation. However, to preserve the spirit of our
approach, we formulate this condition, as we do with our
boundary conditions, for the full theory regardless of
the semiclassical loop expansion. The mass that needs to be taken to
zero is the physical, renormalized, mass scale.  From this point of
view this condition is a trivial consequence of dimensional analysis
if we work with a renormalizable theory where all the physical masses
are proportional to one mass scale. All the masses, including the
soliton mass, vanish then at the same point, which is the conformal,
or critical, point in the theory.  Another way to look at our
condition to fix the integration constant is to consider the Euclidean
version of the 1+1 theory as a classical statistical field theory.
Then the mass of the soliton is the interface tension between two
phases. As is well known \cite{jdl} the interface tension vanishes at
the critical point; moreover, it vanishes with the same
exponent as the inverse correlation length.

In section~\ref{sec:ddm} we demonstrate our new unambiguous method of
calculation using as an example the bosonic kink.  We show that, as
argued in \cite{rvn}, the correct result corresponds to mode number
cutoff.  (The same conclusion was recently reached for nontopological
solitons in 3+1 dimensions \cite{jaffe}.)  In section~\ref{sec:N=1} we
apply our analysis of the topological boundary conditions to the case
of an $N=1$ susy soliton, where it leads to nontrivial
consequences. We analyze the relation of our results to the BPS bound
in section~\ref{sec:bound}. In section~\ref{sec:N=2} we analyze the
$N=2$ susy solitons and conclude that the one-loop corrections vanish
completely.  In section~\ref{sec:2loop} we redo the 2-loop corrections
for the case of the bosonic sine-Gordon soliton \cite{vega,verw},
paying this time close attention to possible ambiguities, and find
that no ultraviolet ambiguities appear.  The ultraviolet ambiguity
 is thus purely a one-loop effect which leads to the
interesting conjecture that it may be formulated in terms of a
topological quantum anomaly.

\section{Eliminating the one-loop ultraviolet ambiguity using
a physical principle}
\label{sec:ddm}

In this section we present a general analysis regarding
the calculation of the soliton mass which will help
us eliminate the ultraviolet ambiguity discussed in \cite{rvn}.
We consider the $\phi^4$ theory (kink) as an example, but the
arguments can be applied to the sine-Gordon theory as well.
The crucial property of these models from which our boundary
conditions follow is the $Z_2$ symmetry $\phi\to-\phi$.

Let us take a step back from the actual calculation and try to 
define the mass of the
soliton {\em before} we do the semiclassical expansion.
We start from the observation that the soliton carries a conserved charge
--- the topological charge.
This means that we can define the mass of the soliton as the difference
between
the energy of the system with nontrivial topology and the energy of the system
with trivial topology. 
This definition coincides with the definition based on path integrals
in the topological sector which are normalized by path integrals
in the trivial sector \cite{vega}.
The topological charge of the system is determined by
the conditions at the spatial boundary. We view the
vacuum energy as a functional
of the boundary conditions. In general, a boundary condition could
induce surface effects associated with the interaction of the
system with the external forces responsible for the
given boundary condition. We would like to avoid these contributions.
There is a class of boundary conditions which do not produce such
effects. 
These are what we call topological boundary conditions, which identify
the degrees of freedom at different points on the boundary
modulo a symmetry transformation. In our case there are two
such possibilities: periodic and antiperiodic. These are  dictated
by the internal $Z(2)$ symmetry: $\phi(x)\to(-1)^p\phi(x)$, $p=0,1$. Crossing
the boundary is associated with a change of variables leaving the
action invariant: $\phi(-L/2)=(-1)^p\phi(L/2)$.  
The system behaves continuously across the
boundary, only our description changes. In effect such boundary
conditions do not introduce a boundary, rather they close
a system in a way similar to the Moebius strip.%
\footnote{Though topological boundary conditions do not induce
boundary effects, finite volume effects vanishing as $L\to\infty$
are of course present. Such effects will be discussed
in section \ref{sec:bound}.
}

We spent so much time on this, perhaps, trivial point in order
to make the choice of boundary conditions for the theory
with fermions clear. The analysis of fermions, however, 
will be postponed until the next section.
In the literature a large number of other boundary conditions have been
considered, both in the bosonic and in the fermionic sectors but from
our perspective they all introduce surface effects or are even
inconsistent.

It should now be clear that the mass of the soliton can be
defined as the difference of the vacuum energy with antiperiodic
and with periodic boundary conditions when the volume $L\to\infty$. 
This definition
does not rely on the semiclassical expansion. Returning
now to the semiclassical calculation we see that at the classical
level the equations of motion select the trivial or the soliton 
vacuum configuration depending on the topology. Less trivially,
we see that at the one-loop level the boundary conditions that
should be used for the small fluctuations about the soliton
configuration should be antiperiodic. We must point out that
the choice of boundary conditions does not affect
the result of the calculation in the purely bosonic case. We
shall nevertheless use antiperiodic boundary conditions
in the soliton sector in this section to be faithful
to our nonperturbative definition of the soliton mass. We shall
see in the next section that for fermions
the choice of the boundary conditions becomes crucial.

A few points about the classical antiperiodic soliton should
be stressed here. The topological boundary condition
reads
\begin{equation}\label{apbc}
\phi(-L/2)=(-1)^p\phi(L/2) \qquad \mbox{ and } \qquad 
\phi^\prime(-L/2)=(-1)^p\phi^\prime(L/2),
\end{equation}
where the nontrivial sector is selected when $p=1$.
Note that the derivative with respect to $x$, $\phi^\prime$,
must also be antiperiodic in the soliton sector.
The classical soliton solution $\phi(x)$ can be viewed
as a trajectory of a particle with coordinate $\phi$ moving
in time $x$ in the potential $-V(\phi)$. The particle
is oscillating about the origin $\phi=0$ with a period which
depends on the amplitude. When the period is equal to $2L$
the trajectory during half of the period is the antiperiodic soliton 
satisfying the boundary conditions (\ref{apbc}). The
endpoints of the trajectory need not necessarily be
the turning points. For example, the particle
at time $-L/2$ can start downhill at some $\phi_0$ with nonzero velocity, 
then pass the point at the same height on the opposite
side, i.e.,  $-\phi_0$, going uphill, then
turn and after that at time $L/2$ pass the point $-\phi_0$ 
again, but going downhill. Clearly, for this trajectory, (\ref{apbc}) is 
satisfied, whereas restricting the usual soliton solution centered at
$x=0$ to the interval
$(-L/2,L/2)$ would lead to a solution for which
$\phi(-L/2)=-\phi(L/2)$, but 
$\phi^\prime(-L/2)=+\phi^\prime(L/2)$.
When $L\to\infty$
the turning points of the trajectory come infinitesimally close
to the minima of $V(\phi)$ and we recover the usual
$L=\infty$ soliton. 

Next, we want to address the problem of the ultraviolet ambiguity
in the one-loop calculation. To summarize the beginning of this
section
\begin{equation}\label{e1-e0}
M \equiv E_1 - E_0,
\end{equation}
where $E_p$, $p=0,1$ are the energies of the system with the periodic
and antiperiodic boundary conditions of (\ref{apbc}) respectively.
At the classical level this gives $M_{\rm cl} \sim
m^3/(3\lambda)$ for the kink, 
where $m$ is the mass of the elementary boson at tree
level, and $\lambda$ the dimensionful coupling constant. 
The order $\hbar$ correction is due to the fact that
boundary conditions change the spectrum of zero-point fluctuations.
Two factors are responsible for the ultraviolet
ambiguity discussed in \cite{rvn}. One is the fact that the terms in the sums
over zero-point energies grow making each sum strongly (quadratically)
ultraviolet cutoff dependent. Second is that one has to compare the
spectrum in two different vacua.  Taking all the modes below a certain
energy in both systems leads to a different result than taking equal
numbers of modes. It would be nice if there was a parameter
in the theory whose variation would continuously interpolate between the two
vacua. This is not possible due to the topological
nature of the difference between the vacua. However, we can
identify a certain value of the dimensionful parameters
of the theory at which the vacuum energy
should become independent of the topology. This will be one ingredient of our
calculation.

Another ingredient is the observation that one can reduce the
ultraviolet divergence of the
sums of zero-point energies by differentiating w.r.t. $m$. 
The terms in the differentiated sums become then decreasing
and as a result the sums (now only logarithmically divergent)
can be unambiguously calculated. But the price
is that we need to supply the value for the integration constant to
recover the function from its derivative.
This can be done using a physical principle that
relates the energies of the two vacua at some value
of the mass. One must
realize that the difference in the energies arises because of the
nontrivial potential for the scalar field. If this potential
vanishes the energies of the two vacua become equal. In the
absence of the potential the mass $m$ of the boson is zero and
the soliton disappears. Therefore
the constant of integration over $m$ is fixed by the condition that the
energy difference between the two vacua must vanish when $m\to0$.
A subtlety here is that $m$ should be sufficiently
large compared to $1/L$ so that finite volume effects can be
neglected. The limit $m\to0$ should be understood in the sense
that the mass approaches $O(1/L)$, where $L$ is large.
Then the difference between the vacuum energies must be $O(1/L)$.
Also note, that other dimensionful parameters in
the theory should be scaled accordingly when $m\to0$, e.g.,
$\lambda/m^2={\rm const}$ in the $\lambda\phi^4$ theory.

We want to relate the mass of the soliton to other parameters of the
theory. The relation to the bare parameters $m_0$ and $\lambda$ 
will contain infinities. The infinities in the
relation of physical quantities to the bare parameters in 
this theory should be eliminated if we renormalize the mass
$m_0 = m + \delta m$, where 
\begin{equation}
\delta m = {3\lambda\over2m}\int_{-\infty}^\infty{dk\over2\pi}{1\over\sqrt{k^2+m^2}}.
\end{equation}
With this renormalization of $m_{0}$ tadpole diagrams vanish. In the
$\phi^4$ theory the
physical pole mass of the meson differs from $m$ by a {\em finite}
amount $-\sqrt3\lambda/(4m)$ \cite{rvn}; however, it suffices to
use $m$ for our purposes. If we rerun this analysis for the
sine-Gordon theory, the tadpole renormalized mass $m$ would,
at one-loop order, coincide with the physical meson mass.
If we use this renormalized mass in the expression for the soliton mass
we get an additional one-loop counterterm, $\delta M$,
\begin{equation}
{m_0^3\over3\lambda} = {m^3\over3\lambda} + \delta M,
\end{equation}
where
\begin{equation}\label{deltaM}
\delta M = {m^2 \delta m\over\lambda} = 
{3m\over2}\int_{-\infty}^\infty{dk\over2\pi}{1\over\sqrt{k^2+m^2}}.
\end{equation}

Now we differentiate the well-known expression for the one-loop 
correction $M^{(1)}$ to the soliton mass with respect to the mass $m$
\begin{equation}\label{M1}
{dM^{(1)}\over dm} = {d(\delta M)\over dm}
	+ {1\over2} \sum_n {d\omega_n\over dm} 
	- {1\over2}\sum_n {d\tilde\omega_n\over dm}.
\end{equation}
For the spectrum $\tilde\omega_n$ in the trivial sector one obtains
\begin{equation}\label{kn_vac}
{d\tilde\omega_n\over dm} = {m\over\sqrt{\tilde k_n^2+m^2}}, \qquad
\tilde k_nL = 2\pi n.
\end{equation}
For the soliton sector
\begin{eqnarray}\label{kn_sol}
{d\omega_n\over dm} = {1\over\sqrt{k_n^2+m^2}}
\left(m + k_n {dk_n\over dm}\right) = 
{1\over\sqrt{k_n^2+m^2}}
\left(m + {1\over L}{k_n^2\over m}\delta^\prime(k_n)\right), 
\nonumber \\
k_nL + \delta(k_n) = 2\pi n + \pi,
\end{eqnarray}
where we used the fact that $\delta(k)$ depends on $m$ only
through $k/m$ to convert the derivative w.r.t. $m$ into the derivative
w.r.t. $k$.

We convert the sums over the spectrum in (\ref{M1}) into integrals
over $k$ using
\begin{equation}\label{sum_triv}
\sum_n f(\tilde k_n) = L \int_{-\infty}^\infty{dk\over2\pi}f(k) + O(1/L);
\end{equation}
and
\begin{equation}\label{sum_delta}
\sum_n f(k_n) = L \int_{-\infty}^\infty{dk\over2\pi}f(k)
	\left(1 + {\delta^\prime(k)\over L}\right) + O(1/L).
\end{equation}
These expressions follow from the Euler-Maclaurin
formula which is valid for a smooth function $f(k)$ vanishing
at $k=\infty$. In our case $f(\tilde k)=d\tilde\omega/dm$ and 
$f(k)=d\omega/dm$ satisfy
these conditions. From the Euler-Maclaurin formula one can also see that 
in the naive calculation with $f(k)=\omega$ the ambiguous 
contribution, which comes from regions
$\delta/L$ at the ultraviolet ends of the integration interval, 
is non-vanishing due to the fact that $f(k)$ grows with $k$.

\begin{figure}[hbt]
\centerline{
\epsfxsize 2in \epsfbox{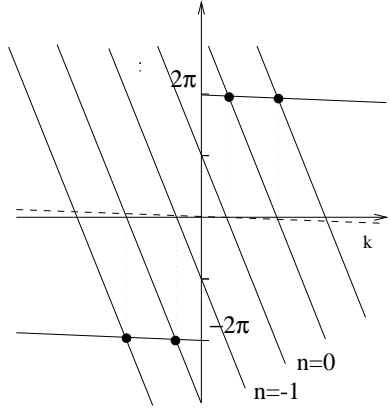}
}
\caption[]{The left- and the right-hand sides of the equation
$\delta(k) = 2\pi n + \pi - kL$ are plotted schematically
by solid lines in the case of the $\phi^4$ kink (two bound states).
The dashed line represents the value of $\delta(k)$ without
the discontinuity $2\pi\varepsilon(k)$. Observe that with this 
discontinuity the mode numbers $n=-1,0$ should be left out,
while the spectrum of allowed values of $k$ is not affected.
}
\label{fig:kink}
\end{figure}

We can use the following expression for the phase shifts $\delta(k)$
in the case of the kink in $\phi^4$ theory:
\begin{equation}\label{deltak}
\delta(k) = \left(2\pi - 2 \arctan {3m|k|\over m^2 - 2 k^2}\right)
\varepsilon(k), 
\end{equation}
where we added the term $2\pi\varepsilon(k)$ to ensure that
$\delta(|k|\to\infty)\to0$. As a result, $\delta(k)$ is discontinuous
at $k=0$: $\delta(0_\pm)=\pm 2\pi$.  It is then easy to see
(Fig.~\ref{fig:kink}) that for
$n=-1$ and $n=0$ the equation (\ref{kn_sol}) does not have solutions
for $k$. It is pleasing to observe that this defect is matched by the
existence of two discrete modes: $\omega_0=0$ (the translational
zero mode) and
$\omega_{-1}=\sqrt3m/2$ (a genuine bound state). To these discrete
modes we can assign (somewhat arbitrarily)
those ``unclaimed'' $n$'s. That this matching is not a coincidence
follows from Levinson's theorem: $\delta(0_\pm)=\pm \pi n_{\rm ds}$,
where $n_{\rm ds}$ is the number of discrete solutions.
Since the discontinuity in $\delta(k)$ is an integer multiple of
$2\pi$, it does not change the spectrum of the allowed values
of $k$ (Fig.~\ref{fig:kink}). This spectrum near the origin is given by
$kL=\ldots,-3\pi,-\pi,\pi,3\pi,\ldots$ up to $O(1/L)$ for
$\delta$ either with or without the $2\pi\varepsilon(k)$ term 
in (\ref{deltak}), and the values of $\delta^\prime(k)$ and $f(k)$
on this set of $k$'s are also not affected.

Putting now all the pieces together we obtain
\begin{equation}
{dM^{(1)}\over dm} = {d(\delta M)\over dm}
	+ {1\over2} \sum_{\rm ds}{d\omega_{\rm ds}\over dm}
	+ {1\over2m}\int_{-\infty}^\infty
	{dk\over2\pi}\sqrt{k^2+m^2}\,\delta^\prime(k).
\end{equation}
This formula is universal.
Substituting for the $\phi^4$ theory the particular values for $\delta M$ 
from (\ref{deltaM}) and
$\delta(k)$ from (\ref{deltak}) we find%
\footnote{Use
$\int_{-\infty}^\infty(1+x^2)^{-1/2}(1+4x^2)^{-1}dx=2\sqrt3\pi/9$
and $\int_{-\infty}^\infty(1+x^2)^{-3/2}dx=2$.
}
\begin{eqnarray}
{dM^{(1)}\over dm} &=& 
-{3m^2\over2}\int_{-\infty}^\infty{dk\over2\pi}{1\over(k^2+m^2)^{3/2}}
	+ {\sqrt3\over4}
\nonumber\\
	&&- {3m^2\over2}\int_{-\infty}^\infty
	{dk\over2\pi}{1\over\sqrt{k^2+m^2}(m^2+4k^2)}
	= {1\over4\sqrt3} - {3\over2\pi}.	
\end{eqnarray}
Integrating over $m$ and using that $M^{(1)}=0$ when $m=0$ 
we obtain the result for the one-loop correction to the 
kink mass which was previously obtained using mode number
cutoff~\cite{dhn1,rvn}
\begin{equation}
M^{(1)} = m\left({1\over4\sqrt3} - {3\over2\pi}\right).
\end{equation}

\section{Fermions, supersymmetry, and topological boundary conditions}
\label{sec:N=1}

In this section we shall extend the ideas introduced in the
previous section to theories with fermions, and 
in particular theories with supersymmetry. The following analysis
can be applied to any $N=(1,1)$ supersymmetric theory with Lagrangian
\begin{equation}\label{lagrangian}
{\cal L} = -{1\over2}\partial_\mu\phi \partial^\mu\phi - {1\over2}U^2(\phi)
- {1\over2}\left( \bar\psi/\hspace{-0.5em}\partial   \psi 
+ U^\prime(\phi)\bar\psi\psi \right),
\end{equation}
where $U(\phi)$ is a symmetric function, admitting a classical
soliton solution. For example, for the kink $U(\phi)=\sqrt{\l/2}(\phi^{2}-
\frac{m_{0}^{2}}{2\l})$. We use $\{\gamma^\mu,\gamma^\nu\}=2g^{\mu\nu}$
with $g^{00}=-1$, $g^{11}=1$, and $\psi$ is a Majorana spinor:
$\bar\psi=\psi^\dagger i\g^0=\psi^TC$ with $C\gamma^\mu=-(\gamma^\mu)^TC$.
The action is invariant under $\d\phi=\bar\epsilon\psi$ and
$\d\psi=(/\hspace{-0.5em}\partial\phi-U)\epsilon$.

First of all we want to identify the class of topological boundary
conditions. The discrete transformation taking $\phi\to-\phi$
must be accompanied by $\psi\to\gamma_3\psi$ with
$\gamma_3=\gamma^0\gamma^1$
to leave the
action invariant. From this
symmetry transformation we obtain topological
boundary conditions
\begin{eqnarray}\label{topbc}
\phi(-L/2) = (-1)^p\phi(L/2),
\qquad 
\phi^\prime (-L/2) = (-1)^p\phi^\prime (L/2),
\nonumber \\
\psi(-L/2) = (\gamma_3)^p\psi(L/2),
\qquad p=0,1.
\end{eqnarray}
The value $p=0$ gives a trivial
periodic vacuum while $p=1$ selects a nontrivial soliton vacuum.  

As one can see, the reasons behind our choice
of the topological boundary conditions do not include supersymmetry.
The same arguments apply to any theory with a Yukawa-like
interaction between fermions and bosons. From this point of view it
is very gratifying to discover that the $p=1$ topological boundary
condition (\ref{topbc})  preserves half of the supersymmetry
of the Lagrangian (\ref{lagrangian}). An easy way to see that
is to consider the Noether current corresponding to the supersymmetry
\begin{equation}
J^\mu = -(/\hspace{-0.5em}\partial\phi + U)\gamma^\mu\psi.
\end{equation}
Integrating 
the conservation equation $\partial_\mu J^\mu=0$ over space we find
\begin{equation}
{\partial Q\over \partial t} \equiv {\partial\over \partial t}
 \int_{-L/2}^{L/2} dx J^0(x)
= -\left[J^1(x)\right]_{-L/2}^{L/2},
\end{equation}
where the r.h.s. is simply the total current flowing into the system.
Using the boundary condition (\ref{topbc}) we obtain
\begin{equation}\label{J1}
-\left[J^1(x)\right]_{-L/2}^{L/2} = 
\left[(/\hspace{-0.5em}\partial\phi + U)\gamma^1\psi 
- (-/\hspace{-0.5em}\partial\phi + U)\gamma^1\gamma^3\psi
\right]_{L/2} = 
\left.(1 + \gamma_3)(/\hspace{-0.5em}\partial\phi + U)\gamma^1\psi 
\right|_{L/2}.
\end{equation}
We see that the $(1-\gamma_3)$ projection of the supercharge $Q$ is
conserved.  Note that different projections of $Q$ are {\em
classically} conserved on the soliton or the antisoliton background.
The soliton with $\phi^\prime+U=0$ and $\psi=0$ is invariant under a
susy transformation with a parameter $\epsilon$ if
$(1+\gamma^1)\epsilon=0$. This means the projection $P_+Q$ (with
$P_\pm=1\pm\gamma_1$) of the
supercharge vanishes on the soliton configuration to linear order in
the quantum fields. For the antisoliton $P_-Q$ has this property.
This should be expected since the topological boundary condition does
not distinguish between the soliton and the antisoliton.

Similarly one can see that the topological boundary
condition does not break translational invariance. The
conservation equation for the stress tensor reads
$\partial_\mu T^{\mu\nu} = 0$, where
\begin{equation}
T^{\mu\nu} = \partial^\mu \phi \partial^\nu \phi 
	+ {1\over4}\left(\bar\psi\gamma^\mu\partial^\nu\psi
	+ \bar\psi\gamma^\nu\partial^\mu\psi \right)
	+ {\cal L}g^{\mu\nu}. 
\end{equation}
In general, the non-conservation of total momentum is again due to the 
boundary term
\begin{equation}
{\partial P\over\partial t} \equiv
{\partial\over\partial t}\int_{-L/2}^{L/2} T^{01} =
-\left[T^{11}\right]_{-L/2}^{L/2}.
\end{equation}
We see that the defining property of the topological
boundary condition, that it relates the fields at
$-L/2$ and $L/2$ by a transformation leaving ${\cal L}$ invariant,
ensures momentum conservation.

Note also that there is another $Z(2)$ symmetry in the Lagrangian
(\ref{lagrangian}): $\psi\to(-1)^q\psi$. This can be used to
extend the set of topological boundary conditions (\ref{topbc}) to
\begin{eqnarray}\label{topbc2}
\phi(-L/2) = (-1)^p\phi(L/2),
\qquad 
\phi^\prime (-L/2) = (-1)^p\phi^\prime (L/2),
\nonumber \\
\psi(-L/2) =  (-1)^q (\gamma_3)^p\psi(L/2),
\qquad p,q=0,1.
\end{eqnarray}
The values $(p,q)=(0,0)$ give a topologically
trivial sector. The sector $(0,1)$
is also trivial, but the fermions have a twist.
Two classically nontrivial vacua are obtained with $p=1$ and $q=0,1$.
For $p=1$ the two values of $q$ correspond to the arbitrariness of the
sign choice of the $\gamma_3$ matrix, and are related to each other
by space parity transformation $\psi(x,t)\to\gamma^0\psi(-x,t)$. 
Therefore one should expect
$E(1,0)=E(1,1)$, which one can check is true at one-loop. 

As in the previous section we define the mass of the soliton as the
difference of the energies $E_p$ of these vacua: $M \equiv E_1 - E_0$.
At the classical level one finds $M=M_{\rm cl}$, where
$M_{\rm cl}$ is the classical soliton mass. The one-loop
correction is determined by integrating
\begin{equation}\label{M1loopSUSY}
{dM^{(1)}\over dm} = {d(\delta M)\over dm} 
+ {1\over2}\sum_n {d\omega^B_n \over dm}
- {1\over2}\sum_n{d\omega^F_n\over dm} - 
\left({1\over2}\sum_n {d\tilde\omega^B_n\over dm} 
- {1\over2}\sum_n{d\tilde\omega^F_n\over dm} \right)
\end{equation}
over $m$. The expressions for the derivatives
$d\omega^B_n/dm$ and $d\tilde\omega^B_n/dm$ are the same as 
in the bosonic case, see (\ref{kn_vac}) and (\ref{kn_sol}).
In order to find the corresponding expressions for the
fermionic frequencies we need to obtain the quantization condition
for $k_n$. For the trivial 
sector we have simply $\tilde k_nL = 2\pi n$. 

The nontrivial sector requires more careful analysis.
The frequencies $\omega$ are obtained by finding solutions of the
equation
\begin{equation}\label{small_ferm}
\left(\gamma^\mu\partial_\mu + U^\prime\right)\psi=0
\end{equation}
of the form $\psi=\psi(x)\exp\{-i\omega t\}$. Multiplying
this equation by $\left(-\gamma^\mu\partial_\mu + U^\prime\right)$
we find
\begin{equation}
\left(-\partial^2 + (U^\prime)^2 + \gamma^1 U^{\prime\prime} U
\right)\psi = 0,
\end{equation}
where we used $\phi^\prime_{\rm sol} = -U(\phi_{\rm sol})$,
which follows from the classical equation of motion
for the soliton in the $L\to\infty$ limit.
Projecting this equation using $P_\pm=(1\pm\gamma^1)/2$
we see that $\psi_+$ (where $\psi_\pm\equiv P_\pm\psi$)
obeys the same equation as the bosonic small fluctuations,
hence
\begin{equation}\label{+as}
\psi_+ = \alpha_+ e^{-i\omega t + i(kx \pm \delta/2)} 
\quad \mbox{ when } x\to\pm \infty.
\end{equation}
The $\psi_-$ component can then be obtained by acting with $P_+$ on
(\ref{small_ferm})
\begin{equation}
\partial_0\gamma^0 \psi_- + \left(\partial_1 + U^\prime\right)\psi_+ = 0.
\end{equation}
which together with (\ref{+as}) gives the asymptotics of $\psi_-$
\begin{equation}\label{thetak}
\psi_- = 
- \gamma^0 \alpha_+{k\mp im\over\sqrt{k^2+m^2}}
e^{-i\omega t + i(kx \pm \delta/2 )} = 
- \gamma^0 
\alpha_+ e^{\pm i\theta/2} e^{-i\omega t + i(kx \pm \delta/2 )} 
\quad \mbox{ when } x\to\pm \infty,
\end{equation}
where $\theta(k) = -2\arctan(m/k)$ and we used the fact that
in the $L\to\infty$ limit $U^\prime\to\pm m$.
Therefore the solutions of (\ref{small_ferm})
have asymptotics
\begin{equation}\label{evectorgamma}
\psi = \psi_+ + \psi_- = \left(
1 - \gamma^0 e^{\pm i\theta/2}\right)\alpha_+ 
e^{-i\omega t + i(kx \pm \delta/2)}
\quad \mbox{ when } x\to\pm \infty
,
\end{equation}
where $\alpha_+$ is the eigenvector of $\gamma^1$ with eigenvalue $+1$.

Although one could continue the derivation without specifying the
representation for the $\gamma$ matrices (an exercise for the reader)
we find it more concise to adopt a certain representation.  The most
convenient is the following one in terms of the
Pauli matrices: $\gamma^0 = -i\tau_2$,
$\gamma^1=\tau_3$, and hence $\gamma_3=\tau_1$. 
It has two advantages. First,
$\alpha_+$ has now only an upper component. Second, 
the Majorana condition becomes simply $\psi^*=\psi$
and the equation (\ref{small_ferm}) is real.
In this representation (\ref{evectorgamma}) becomes
\begin{equation}\label{as_psi}
\psi = \left( \begin{array}{c}
1  \\ - e^{\pm i\theta/2}
\end{array}\right) \alpha e^{-i\omega t + i(kx \pm \delta/2)}
\quad \mbox{ when } x\to\pm \infty,
\end{equation}
where $\alpha$ is a complex number.

Now we impose the boundary condition
$\psi(-L/2) = \gamma_3\psi(L/2)$. The field $\psi$ in equation
(\ref{small_ferm}) must be real. This means that
only the real part of (\ref{as_psi}) need to satisfy the boundary
condition. This condition should, however, be maintained for
all $t$. Therefore, due to the oscillating phase $\exp(-i\omega t)$,
a complex equation must be satisfied
\begin{equation}\label{BCGamma}
e^{-i(kL+\delta)} 
\left( \begin{array}{c} 1 \\ - e^{ -i\theta/2}  \end{array}\right)
=\Gamma
\left(\begin{array}{c} 1 \\ - e^{ i\theta/2} \end{array}\right)
\end{equation}
We introduced $\Gamma=\gamma_3$ in order to discuss briefly
the following point. One could consider a more general
boundary condition: $\psi(-L/2) = \Gamma\psi(L/2)$ with some matrix
$\Gamma$ (real in the Majorana representation we have chosen).
One can see from (\ref{BCGamma}) that certain boundary
conditions cannot be satisfied, for example, 
the frequently employed \cite{kar1,CM,Uch,uchi1,uchi2,rvn}
periodic boundary conditions with $\Gamma=1$.
Equation  (\ref{BCGamma}) provides an additional consistency
check for our choice of boundary condition.

With  the topological boundary condition $\Gamma=\gamma_3=\tau_1$,
we find that equation (\ref{BCGamma}) is satisfied provided
\begin{equation}\label{kn_ferm}
kL + \delta + {\theta\over2} = 2 \pi n + \pi.
\end{equation}
Using this quantization rule we find
\begin{equation}\label{doF/dm}
{d\omega^F_n\over dm} = {1\over\sqrt{k^2+m^2}}
\left(m + k_n {dk_n\over dm}\right) = 
{1\over\sqrt{k^2+m^2}}\left(m 
+ {1\over L}{k_n^2\over m}
\left(\delta^\prime(k)+{\theta^\prime(k)\over2}\right)\right)
\end{equation}

\begin{figure}[hbt]
\centerline{
\epsfxsize 2in \epsfbox{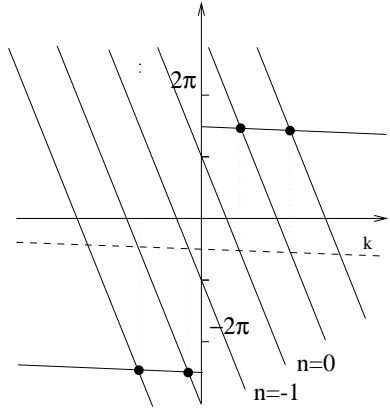}
}
\caption[]{
The left- and the right-hand sides of the equation $\delta
+ \theta/2 = 2\pi n + \pi - kL$ are plotted schematically in the case
of the supersymmetric $\phi^4$ kink (two bound states).  The dashed
line represent the value of $\delta(k)+\theta(k)/2$ without the discontinuity
$2\pi\varepsilon(k)$. As in the bosonic spectrum the discontinuity
leads to $n=-1,0$ mode numbers being skipped, while the spectrum of
allowed values of $k$ is not affected.  
}
\label{fig:skink}
\end{figure}

Now we convert the sums over modes into integrals.
For the bosonic and fermionic sums in the trivial sector
 formula (\ref{sum_triv}) applies and the
sums cancel each other (no cosmological constant in the 
trivial susy vacuum). For the bosonic sum
in the nontrivial sector we use again (\ref{sum_delta}).
For the fermionic sum a formula analogous to (\ref{sum_delta})
applies with $\delta+\theta/2$ instead of $\delta$, which
follows from (\ref{kn_ferm}). Again, as for the bosonic modes,
due to the discontinuity in $\delta$ there are $n_{\rm ds}$
values of $n$ which do not lead to a solution of (\ref{kn_ferm}).
The remaining $n$ lead to $k$ values (namely, in the case of the kink:
$kL=\ldots,-5\pi/2,-\pi/2,3\pi/2,7\pi/2,\ldots$, see
Fig.~\ref{fig:skink}) which in the
limit $L\to\infty$ give a continuous integration measure
$Ldk/(2\pi)$ near $k=0$.

Putting now everything into  formula (\ref{M1loopSUSY})
we find
\begin{equation}\label{dm1theta}
{dM^{(1)}\over dm} = {d(\delta M)\over dm}
-{1\over4m}\int_{-\infty}^\infty {dk\over2\pi} \sqrt{k^2+m^2}\,\theta^\prime(k).
\end{equation}
The one-loop mass counterterm is given by \cite{rvn}
\begin{equation}\label{deltaMSUSY}
\delta M = {m\over2} \int_{-\infty}^\infty {dk\over2\pi} {1\over\sqrt{k^2+m^2}}.
\end{equation}
It follows from the renormalization counterterm $\delta m$ which is
chosen to cancel the sum of the bosonic and fermionic tadpole
diagrams.  Substituting into (\ref{dm1theta}) we find
$dM^{(1)}/dm=-1/(2\pi)$, hence
\begin{equation}\label{M1SUSY}
M^{(1)} = - {m\over2\pi}.
\end{equation}

This result differs from the one two of us have obtained previously
\cite{rvn} using a mode-number regularization scheme with the
conventionally employed (but, as we have argued, untenable) periodic
boundary conditions.

In the special case of the supersymmetric sine-Gordon model,
we can compare this result with the one  obtained from the Yang-Baxter
equation assuming the factorization of the S-matrix \cite{scho}.
The mass spectrum is then given by~\cite{ahn}
\begin{equation}\label{m_n}
m_n = 2M\sin(n\gamma/16),
\end{equation}
where $\gamma$ in the notation of ref. \cite{ahn} is related to
the bare coupling $\beta$ through
\begin{equation}\label{gammabeta}
{1\over\gamma} = {1 - \beta^2/4\pi\over4\beta^2} 
	= {1\over4\beta^2} - {1\over16\pi}.
\end{equation}
Expanding (\ref{m_n}) for $n=1$ we find
\begin{equation}
m_1 = {M\gamma\over8} + O(\gamma^3).
\end{equation}
Since this is the lightest mass in the spectrum we identify
it with the meson mass $m=m_1$. Taking the ratio $M/m_1$ and using
(\ref{gammabeta}) we obtain
\begin{equation}
{M\over m} = {8\over\gamma} + O(\gamma) = {2\over\beta^2}
	- {1\over2\pi} + O(\beta^2).
\end{equation}
The first term is the classical result, the second is the 1-loop
correction. This means that the 1-loop correction to $M$ following
from the exact S-matrix factorization calculation
\cite{ahn} is the same as our 1-loop result (\ref{M1SUSY}).

The next question we address is whether such a negative
correction is in agreement with the well-known BPS bound. 
This question has been subject to controversy and deserves a separate
section.


\section{Quantum BPS bound, soliton mass, and finite size effects}
\label{sec:bound}

As was first realized by Olive and Witten
\cite{WO}, the naive supersymmetry algebra in a topologically
nontrivial sector is modified by central charges.  The susy generators
for the $N=1$ model read in the representation of section 3
\begin{equation}\label{qpm}
Q_\pm\equiv P_\pm\int_{-L/2}^{L/2} [ - 
(/\hspace{-0.5em}\partial\varphi + U)\gamma^0\psi ] dx
=\int_{-L/2}^{L/2} [ \dot\varphi\psi_\pm+(\varphi'\pm U)\psi_\mp]dx,
\end{equation}
where $P_\pm$ is again $(1\pm\gamma_1)/2$. Using canonical commutation
relations we arrive at the following algebra
\begin{equation}\label{algebra}
\left\{ Q_\pm, Q_\pm \right\}=2H\pm 2Z;\quad
\left\{ Q_+,Q_-\right\}=2P,
\end{equation}
where
\begin{eqnarray}\label{hpz}
H&\equiv&\int_{-L/2}^{L/2} [ \frac12 \dot\varphi^2+\frac12(\varphi')^2+
\frac12 U^2
+ \frac{i}2(\psi_+\psi_-'+\psi_-\psi_+')-iU'\psi_+\psi_-]dx \\
P&\equiv&\int_{-L/2}^{L/2} [ \dot\varphi\varphi'+
\frac{i}2(\psi_+\psi_+'+\psi_-\psi_-')]dx \\
Z&\equiv&\int_{-L/2}^{L/2} \varphi' U dx
\end{eqnarray}
The central charge $Z$ is clearly a boundary term.

Let us comment on some subtleties in the derivation of
(\ref{algebra}). The Dirac delta 
functions in the equal-time canonical commutation relations can be written as
$\delta (x,y)=\sum_{m} \eta_{m}(x)\eta_{m}(y)$, where $\eta_{m}$ is  a complete
set of functions {\em satisfying the boundary conditions} of the corresponding 
field. For such $\delta (x,y)$ one has:
\begin{eqnarray}
\int_{-L/2}^{L/2}\phi(x)\delta(x-L/2)&=&\frac{\phi(L/2)\pm\phi(-L/2)}{2}\\
\int_{-L/2}^{L/2}\phi(x)\delta(x+L/2)&=&\frac{\phi(-L/2)\pm\phi(L/2)}{2}
\end{eqnarray}
For the bosons in the topological sector we need the $-$ signs. For the 
fermions $\psi_{+}(x)+\psi_{-}(x)$ one needs the $+$ signs, 
but for the fermions
$\psi_{+}(x)-\psi_{-}(x)$ one needs the $-$ signs. That {\em some} subtlety in 
the delta functions is present, is immediately clear if one considers the 
double integral $\int\int dxdy\;f(x)\partial_{x}\delta(x-y)g(y)$, and either 
directly partially integrates the derivative $\partial/\partial {x}$, or first 
replaces $\partial_{x}\delta(x-y)$ by $-\partial_{y}\delta(x-y)$ and then 
partially integrates w.r.t. $y$. One gets the same result provided
\begin{equation}
fg(x)|_{x\in B}=\int_{-L/2}^{L/2}dy\;f(x)\delta(x-y)g(y)|_{x\in B}+
\int_{-L/2}^{L/2}dx \;f(x)\delta(x-y)g(y)|_{y\in B}
\label{consist}
\end{equation}
where the notation $h(x)|_{x\in B}$ implies $h(L/2)-h(-L/2)$. Naively, there 
is a factor of 2 missing in this equation, but with the more careful 
definitions of $\delta(x-y)$ for periodic or antiperiodic functions, 
consistency is obtained.

With these delta functions one finds that the boundary terms in the 
$\{ Q_{\pm}, Q_{\pm}\}$ anticommutators reproduce the consistency condition
(\ref{consist}), hence cancel, whereas in the $\{Q_{+},Q_{-}\}$ anticommutator 
one finds the boundary term 
\begin{equation}
-i\int dx\psi_{+}(x)\delta_{a}(x-y)\psi_{+}(y)|_{y\in B}-i\int dx \psi_{-}(x)
\delta_{a}(x-y)\psi_{-}(y)|_{y\in B}
\end{equation}
where $\delta_{a}(x-y)$ is the bosonic (antisymmetric) delta function defined
above. These terms cancel if one uses our topological boundary conditions.
In the $\{ Q,Q\}$ relations one does not encounter subtleties involving delta
functions for the fermions because there are no derivatives of fermions in 
$Q_{\pm}$.

Since the operators $Q_\pm$ are
hermitian, one finds that the following relation exists between the
expectation values of operators $H$ and $Z$:
\begin{equation}\label{thebound}
\langle s | H | s \rangle \geq | \langle s | Z | s \rangle |,
\end{equation}
for any quantum state $s$.

As we have already pointed out in section 3 (\ref{J1}), only one linear
combination of $Q_+$ and $Q_-$ is conserved in the soliton sector:
$Q_+\pm Q_-$ for $q=1,0$ respectively. Taking for definiteness $q=0$
we derive from (\ref{qpm}),(\ref{hpz}) the following commutation relations:
\begin{eqnarray}\label{47}
i[H,Q_+ + Q_-]&=&
2[(\dot\varphi + \varphi')(\psi_+ +\psi_-)+U(\psi_+ - \psi_-)
]_{x=L/2} \\ 
i[P,Q_+ + Q_-]&=&
2[(\dot\varphi + \varphi')(\psi_+ +\psi_-)-U(\psi_+ - \psi_-)
]_{x=L/2},
\end{eqnarray}
while the other linear combination $Q_+-Q_-$ commutes with both $H$
and $P$.\footnote{In these relations one must use
the proper definitions of the delta functions for the
fermions since derivatives of $\psi_\pm$ appear in $H$.
As one can see,  (\ref{47}) agrees with (\ref{J1}).}
We also find that the operator $Z$ does not commute
with the Hamiltonian in the soliton sector:
\begin{equation}
i[H,Z] = 2U\dot\varphi\Big|_{x=L/2}
\end{equation}

Let us examine carefully the meaning of this last result.  Strictly
speaking, it implies that $Z$ is not a central charge. We shall show
that this fact reflects a certain property of $Z$ in a {\it finite}
volume $L$.  Let us ask the following question: what is the
expectation value of $Z$ in the soliton vacuum state $|{\rm
sol}\rangle$?  It is clear that for any finite volume $L$: $\langle
{\rm sol} | Z | {\rm sol}\rangle =0$, since neither a positive nor a
negative value is distinguished by our boundary conditions. This is a
consequence of the fact that $H$ and $Z$ do not commute. The ground
state $|{\rm sol}\rangle $ is an eigenvector of $H$, but it need not
be and is not an eigenvector of $Z$.

To make the next step clear it is convenient to use the following
observation: The value of $Z$ measures the position of the
soliton. Indeed, for the classical configuration if the center of the
soliton is at $x=0$ exactly then $Z$ is $M$ (up to $O(1/L)$). If we
now move the center of the classical soliton the value of $Z$ will
decrease and reach zero when the center of the soliton is at $L/2$. If
we continue shifting the solution in the same direction, and bearing
in mind its antiperiodicity, $Z$ will become increasingly negative
and it will reach the value $-M$ when there is an antisoliton at
$x=0$. If we deal with a configuration which is a distortion of the
soliton, the center of the soliton is not well defined but $Z$ is and
can give one an idea of where the soliton is. On the quantum level
this corresponds to the fact that $P$ and $Z$ do not commute.  If we
act with $P$ (this is our spatial shift) on an eigenstate of $Z$ we
generate a different eigenstate of $Z$ and the expectation value of
$Z$ changes.

Now, if we think of $Z$ as (a nonlinear function of) the coordinate
and $P$ as the momentum of the soliton, the next step becomes
clear. The vacuum is an eigenstate of $P$ with eigenvalue
0. Therefore, it is a superposition of the eigenstates of
$Z$. Positive and negative values of $Z$ enter with the same weight
into this superposition (the corresponding eigenvectors can be
obtained from each other by acting with $\exp(iLP)$ --- a shift by
$L$). Therefore $\langle {\rm sol} | Z | {\rm sol} \rangle$=0 for any
finite $L$.

It is, however, too early to conclude that the equation
(\ref{thebound}) does not lead to any condition on $\langle {\rm
sol}|H|{\rm sol}\rangle$ apart from semipositivity. The expectation
value of $Z$ can be compared to an order parameter in a system with
spontaneous symmetry breaking. It is only nonzero if the thermodynamic
limit $L\to\infty$ is taken properly. We shall now analyze how the
limit $L\to\infty$ must be taken in the case of the soliton.

One can view the soliton as an almost classical particle (as long as
the coupling constant is small) which is subject to Brownian motion
due to quantum fluctuations. This is the meaning of the fact that $H$
and $Z$ do not commute: $Z$, or the position of the soliton, depends
on time. If we wait for a sufficiently long time it will cover all
possible positions and the expectation value of $Z$ will be
zero. However, it is obvious that for large $L$ most of the time the
soliton will spend away from the boundaries.  It means that if one
starts from a state of the soliton away from the boundary so that $Z=M$
and limits the time of observation the expectation value of $Z$
will remain close to $M$.

How long does it take for the soliton to cover all the volume $L$?
Since it is a random walk the distance from the original position is
$O(\sqrt t)$ and it will stay away from the boundary if we restrict
the interval $t$ of the observation to $t\ll O(L^2)$. Such a
restriction will mean that we introduce an error of at most $O(1/L^2)$
in the energy due to uncertainty principle.  This is small when $L$ is
large. Alternatively, one can do Euclidean rotation and consider
classical statistical theory of an interface in 2 dimensions. The
random walk in this case is the well-known roughening of the interface
\cite{jdl}. It leads to smearing of the interface to the width of
$w={\rm const}\sqrt{t/M}$, where $M$ is the one-dimensional interface
tension, which is the soliton mass. The correction to the tension
turns out to be $O(1/t)$, as we have already seen. (More rigorously,
the partition function of the wall is not just $\exp(-Mt)$ in the
$t\to\infty$ limit but has a preexponent $L\sqrt{M/t}$ due to the
fluctuations of the interface. The factor $L$ arises from the integration
over the volume of the collective coordinate with a familiar measure
$\sqrt{Mt}$, and an additional factor $1/t$ comes from
the determinant of nonzero soft vibrational modes of the interface 
\cite{amst}).

Therefore the thermodynamic limit $L\to\infty$ in our system which
gives nonzero $\langle {\rm sol}|Z|{\rm sol} \rangle$ corresponds to
the energy measurement whose duration $t$ is small compared to $ML^2$.
The bound (\ref{thebound}) must apply to the result of such a
measurement.  The expectation value of $Z$ in such a thermodynamic
limit is what was calculated in \cite{imb1,rvn}:
\begin{eqnarray}
Z &=& Z_{\rm cl}(m_0) + \int_{-\infty}^{\infty} dx {d\over dx}
\left[\frac12 U^\prime(\phi_{\rm sol}(x))\langle \eta^2(x)\rangle\right]
\nonumber \\
&=& M_{\rm cl}(m_0) - \frac m2 \int_{-\infty}^{\infty} {dk\over2\pi}
{1\over\sqrt{k^2+m^2}} = M_{\rm cl}(m).
\end{eqnarray}
The logarithmically divergent integral is exactly cancelled by the
counterterm $\delta M$ (\ref{deltaMSUSY}) after the renormalization
of $m$.

The question we need to answer now is what is the value of 
$\langle {\rm sol}|H|{\rm sol} \rangle$? We recall that our calculation
of the soliton mass was aimed at finding the dependence of
$M$, or, $\langle {\rm sol}|H|{\rm sol} \rangle$ on $m$, the renormalized mass.
We were not able to determine a constant term in the
unrenormalized $\langle {\rm sol}|H|{\rm sol} \rangle$, but we knew that it
must be subtracted to satisfy the renormalization condition
$M|_{m=0}=0$. To evaluate the l.h.s. of (\ref{thebound})
we have to know this constant. In order to find it
we must evaluate directly the sums of bosonic and fermionic frequencies
in the soliton sector. Although the sums are quadratically
divergent supersymmetry improves  the situation. It
requires that bosonic and fermionic modes come
in pairs. Therefore we need to apply the Euler-Maclaurin formula
to a function $f(n)=(\omega_n^B-\omega_n^F)/2$ which has much better
behavior at large $n$. Using the spectral relations for the bosons 
(\ref{kn_sol}) and the fermions (\ref{kn_ferm}) we find:
\begin{eqnarray}
&\langle {\rm sol}|H|{\rm sol} \rangle &= M_{\rm cl}(m_0)
+ \frac12 \sum_{n=-N}^{N} (\omega_n^B-\omega_n^F)
=
M_{\rm cl}(m_0) + \frac12 \int_{-\Lambda}^{\Lambda} {dk\over2\pi}
{\theta(k)\over2}{d\over dk}\sqrt{k^2+m^2}
\nonumber\\
&&=
M_{\rm cl}(m_0) +
{1\over8\pi}\sqrt{k^2+m^2}\,\theta(k)\Big|_{-\Lambda}^{\Lambda}
- \frac14 \int_{-\Lambda}^{\Lambda}{dk\over2\pi}
\sqrt{k^2+m^2}\,\theta^\prime(k) 
\nonumber\\
&&\stackrel{\Lambda\to\infty}=
M_{\rm cl}(m_0) +
{\Lambda\over4} - {m\over 2\pi}
- \frac12 \int_{-\infty}^{\infty}{dk\over2\pi}{m\over\sqrt{k^2+m^2}}
.
\end{eqnarray}
The constant $\Lambda$ is the ultraviolet cutoff related to the number
of modes: $L\Lambda=2\pi (N+1/2) + O(1/L)$, and we recall that
$\theta=-2\arctan(m/k)$. The last integral is logarithmically
divergent and is exactly cancelled by the counter-term $\delta M$
(\ref{deltaMSUSY}) when we renormalize $m_0=m+\delta m$. The same
divergence appears in $\langle {\rm sol}|Z|{\rm sol} \rangle$ and is
also removed when $m$ is renormalized.  On the other hand, the linear
divergent term $\Lambda/4$ does not appear in $\langle {\rm
sol}|Z|{\rm sol} \rangle$. In terms of the renormalized mass $m$ and
$M_{\rm cl}(m)$ the left- and the right-hand sides of (\ref{thebound})
are:
\begin{equation}
\langle {\rm sol}|H|{\rm sol} \rangle = M_{\rm cl} + {\Lambda\over4} - {m\over2\pi}
\qquad \mbox{ and } \qquad \langle {\rm sol}|Z|{\rm sol} \rangle = M_{\rm cl}. 
\end{equation}
We see that the bound is observed by the soliton vacuum state with an
infinite overkill due to a linearly divergent constant $\Lambda/4$. This
constant is only nonzero in the soliton sector. In
this sector the bound is not saturated, which shows that ``what can
happen --- does happen'': the argument of Olive and Witten for the
saturation of the bound is based on the ``multiplet shortening'' and
does not apply to $N=1$ susy solitons.

In other words, to resolve the long-standing problem of the Bogomolnyi
bound in the $N=1$ supersymmetric soliton/kink model we must realize
that the bound is imposed on the unrenormalized expectation value of
the Hamiltonian $\langle {\rm sol}|H| {\rm sol} \rangle$.  Both sides
of equation (\ref{thebound}) contain ultraviolet divergences.  The
supersymmetry ensures that quadratic divergences do not appear in
$\langle H \rangle$.  However, $N=1$ is not enough and a linear
divergence remains in the topologically nontrivial sector (there is
also a logarithmic divergence, but it is matched on both sides,
$\langle H \rangle$ and $\langle Z \rangle$).  This linearly divergent
term is positive in accordance with the bound (\ref{thebound}).
Note that this divergence is different from a cosmological constant
(which vanishes because of supersymmetry) in that it is proportional
to $L^0$ rather than $L^1$.

Since $\langle {\rm sol}|H|{\rm sol} \rangle$ is divergent even after
standard renormalization of the mass $m$ we need to use an additional
renormalization condition to find the physical soliton mass. This is
the condition $M|_{m=0}=0$ which we introduced.  Therefore
\begin{equation}
M=\langle {\rm sol}|H|{\rm sol} \rangle - \langle {\rm sol}|H|{\rm sol} \rangle|_{m=0} = 
\langle {\rm sol}|H|{\rm sol} \rangle - {\Lambda\over4}.
\end{equation}
Our new renormalization condition is based on the physical requirement
that the physical vacuum energy should not depend on topology in the conformal
point $m=0$. Therefore, in principle, it also requires a subtraction
of the expectation value of $H$ in the topologically trivial vacuum
$\langle0|H|0\rangle$, or rather $\langle0|H|0\rangle -
\langle0|H|0\rangle|_{m=0}$. In the non-supersymmetric case this is
essential to cancel the background bulk contributions linear in $L$,
which are $m$-dependent, and which are the same in both sectors.
However, supersymmetry ensures that $\langle0|H|0\rangle=0$. Finally,
we find a negative finite quantum correction to the physical soliton
mass given by (\ref{M1SUSY}).  However, it is not the physical mass
$M$, but rather it is $\langle {\rm sol}|H|{\rm sol} \rangle$, to
which the bound (\ref{thebound}) applies. As we shall see in the next
section, in the case of $N=2$ supersymmetry all corrections, even
finite, are cancelled and
$M=\langle {\rm sol}|H|{\rm sol} \rangle$ saturates the bound as the
``multiplet shortening'' demands.

\section{The N=2 case}
\label{sec:N=2}

Consider the action for the following N=(2,2) susy model in
1+1 dimensions
 \eq\label{n2action}
 \cl = - \del_\m \varphi^* \del^\m \varphi - \bar{\j} \g^\m
\del_\m \j - U^* U - {1\over 2} (U' \j^T i \g^0 \j + U^{*'} \j^{*T}
i \g^0 \j^*) 
\eqe
where $\varphi$ and $\j$ are complex, $U=U (\varphi ), U' =
(\del/\del \varphi) U$, and $\g^1 = \t_3$ and $\g^0 =- i \t_2$
with $\t_3$  and $\t_2$ the usual Pauli matrices.\footnote{With
$\g^1=\t_3$ instead of $\g^1 = \t_1$ the diagonalization of the
fermionic actions is easier.}  
The action is invariant under
$\d \varphi = \bar{\e} \j, \d \varphi^* = \bar{\j} \e, \d \j = 
/\hspace{-.5em}\partial
\varphi \e - U^* \e^*$ and $\d \bar{\j} = - \bar{\e} 
/\hspace{-.5em}\partial \varphi^* - U \e^T i
\g^0$ with complex two-component spinors
$\e$, with $\bar{\e}=\e^\dag i\g^0$ and $\bar{\j}=\j^\dag i\g^0$.  

In terms of the components $\varphi = (\varphi_1 + i
\varphi_2)/\sqrt{2}$ and $\j^T = (\j_+, \j_-)$ the action reads
 \eqa
 \cl &=& - \half \del_\m \varphi_1 \del^\m \varphi_1 - \half
\del_\m \varphi_2 \del^\m \varphi_2 + i \j_+^* (\dot{\j_+} -
\j_-^\prime) + i \j_-^* (\dot{\j}_- - \j_+^\prime)
\nonumber\\
&& - U^*U +  i U' \j_+ \j_- +  i {U^*}'
\j_+^*  \j_-^*
\eqae
For the superkink 
\eq
U_{\mathrm kink}=\sqrt\l(\varphi^2-\mu_0^2/2\l)
\eqe
while for
super sine-Gordon theory
\eq
U_{\mathrm sine-Gordon}=m_0^2\sqrt{2/\l}\cos(\sqrt{\l/2}\phi/m_0)
\eqe
In terms of $\varphi_1$ and $\varphi_2$ the potential is given by
\eqa
U^*U_{\mathrm kink}&=&
{\l \over 4} \left[ \left( \varphi_1^2 - \m_0^2 /\l \right)^2
+ 2 \varphi_1^2 \varphi_2^2 + \varphi_2^4 + 2
{\m_0^2 \over \l} \varphi^2_2 \right]\\
U^*U_{\mathrm sine-Gordon}&=&{m_0^4\over\l}\left[
\cos{\sqrt\l\over m_0}\varphi_1+\cosh{\sqrt\l\over m_0}\varphi_2\right]
\eqae
This already shows that for the kink 
the trivial solutions  $\varphi^{(0)}_1 = \pm
\m_0/\sqrt{\l}$ and the kink solution $\varphi_1=\varphi_K = 
( \m_0/\sqrt{\l}) \tanh\m_0 x/\sqrt{2} $ 
(and the antikink solution
$\varphi_{\bar{K}} = - \varphi_K)$ of the bosonic kink model
remain solutions of this susy model
while $\varphi_2=0$.  Because the potential is of
the form $(\varphi^2 - \m^2_0 /2\l) (\varphi^{*2} -
\m^2_0/2\l)$ instead of $(\varphi \varphi^* - \m_0^2 / 2\l)^2$,
there is no U(1) symmetry acting on $(\varphi_1 , \varphi_2)$
which can rotate the kink away.  Hence there is a genuine
soliton.  

In sine-Gordon theory, the trivial vacua are at $\varphi^{(0)}_2=0$ and
$\varphi^{(0)}_1=(n+{1\over2})\pi m/\sqrt\l$, while for the vacuum in the
topological sector we choose the solution $$\varphi^{\mathrm sG}_{\mathrm
sol}(x)=m/\sqrt\l(4\arctan(\exp mx)-\pi).$$ This solution is
antisymmetric in $x$, in agreement with the $Z_2$ topological boundary
conditions of section~2.\footnote{This solution
is obtained from the one in \cite{rvn} by the substitution
$\varphi=\varphi'+m\pi/\sqrt\l$. Actually, the sine-Gordon model has $Z$
symmetry, and we could choose solutions which interpolate between two
other minima of the potential by constant shifts. }

The transformation rules in component form read
 \eqa
\d \varphi = - i \e_+^* \j_- + i \e_-^* \j_+ && \d \j_+ =
\varphi^\prime \e_+ - \dot{\varphi} \e_- - U^* \e^*_+
\nonumber\\
\d \varphi^* = - i \j_+^* \e_- + i \j_-^* \e_+ && \d \j_- = -
\varphi^\prime \e_- + \dot{\varphi} \e_+ - U^* \e_-^*
\eqae
and since for the soliton 
$\varphi_{\mathrm sol}^\prime + \sqrt{2}U (\varphi_{\mathrm sol}/\sqrt2) =
0$, the solution $\varphi = (1/\sqrt{2}) \varphi_{\mathrm sol} , \j_+ = \j_- =
0$ is only preserved by half the susy transformations, namely
those with $ {Im}\; \e_+$ and $ {Re}\; \e_-$.

In the sector without soliton we set $\varphi_1 = \varphi^{(0)}_1 + \h$ 
and find then the following linearized field
equations
 \eqa
&& \Box \h - m_0^2 \h = 0 \; , \; \Box \varphi_2  - 
m_0^2 \varphi_2 = 0 \nonumber\\
&& \left. \begin{array}{l} \dot{\j}_+ - \j_-^\prime +
m_0 \j_-^* = 0 \\ \dot{\j}_- - \j_+^\prime - m_0
\j_+^* = 0  \end{array}\right\}  \begin{array}{l} \ddot{\j}_+
-\j_+^{\prime\prime} + m_0^2 \j_+ = 0 \\ \ddot{\j}_- -
\j_-^{\prime\prime} + m_0^2 \j_- = 0 \end{array}
\eqae
where $m_0=U'(\phi_1^{(0)}/\sqrt2)$ and we used that
$U(\phi_1^{(0)}/\sqrt2)=0$.
Hence all fields satisfy the same second-order field equation
in the trivial sector, with a common mass $m_0$ (for the kink $m_0^2 = 2
\m_0^2$).

In the sector with a soliton, we set $\varphi_1 = \varphi_{\mathrm sol} (x)
+ \h$, and find then the linearized field equations
\eqa
\label{lfe}
 \Box \h - (U'^2+UU'')\Big|_{\varphi_{\mathrm sol}} \h=0 &;&
\dot{\j}_+ - \j_-^\prime + U'\Big|_{\varphi_{\mathrm sol}} \j_-^* = 0
\nonumber\\
 \Box \varphi_2 - (U'^2-UU'')\Big|_{\varphi_{\mathrm sol}}
\varphi_2 = 0 &;& \dot{\j}_- - \j_+^\prime - U'\Big|_{\varphi_{\mathrm sol}}
 \j_+^* = 0
\eqae
Decomposing $\j_+ = { Re}\, \j_+ + i { Im}\, \j_+$ 
and similarly for $\j_-$, 
the fermionic field equations split into one pair of equations
which couple $Re\, \j_+ $ to   $Re\, \j_-$, and another pair which
couple $Im\, \j_+$ to $Im\, \j_-$.  Iteration and the relation
$\varphi_{\mathrm sol}'=-\sqrt2 U$ lead to
 \eqa
&& Re\, \j_+^{''} - Re\, \ddot{\j}_+ - (U'^2+UU'')\Big|_{\varphi_{\mathrm sol}}
 Re\, \j_+ = 0 \; , \; {\rm idem \; for } \; Im\, \j_-
\nonumber\\
&& Re\, \j_-^{''} - Re\, \ddot{\j}_- - (U'^2-UU'')\Big|_{\varphi_{\mathrm sol}}
 Re\, \j_- = 0 \; , \; {\rm idem \; for} \; Im\, \j_+
\eqae
Hence, the real triplet $\h, Re\, \j_+$ and $Im\, \j_-$ satisfies
the same field equation as the real scalar $\h$ and the real upper
component $\j_+$ of the Majorana fermion in the N=(1,1)
model, whereas the real triplet $\varphi_2, Im\, \j_+$ and $Re\,
\j_-$ satisfies the same field equation as the real spinor
$\j_-$ in the N=(1,1) model \cite{rvn}.  

In principle one can directly determine the discrete spectrum
of $\varphi_2$ by solving the Schr\"odinger equation with
$V=(U')^2-UU''$ \cite{mor}, but susy already gives the
answer. In the N=(1,1) model, the spinors $u_\pm $ in $\j_\pm
(x,t) = u_\pm (x) \exp - i \omega t$ satisfy the coupled equations
$(\del_x + \tilde{U}^\prime) u_+ + i \omega u_- = 0$ and $(\del_x -
\tilde{U}^\prime) u_- + i \omega u_+ = 0$, where $\tilde{U}=
\sqrt2 U(\varphi=\varphi_1/\sqrt2)$. 
Any solution for $u_{+}$ and $u_{-}$ yields 
also a solution for $\varphi_2$ (with $\varphi_2 \sim u_{-}$),
and any solution for $\varphi_2$ leads also to a solution for $u_{+}$ and 
$u_{-}$ (with $u_{-}\sim\varphi_2$ and $u_{+}\sim (\del_{x}-\tilde U^\prime)
\varphi_2$).  Hence, there are as many 
bound states for $\varphi_2$ as for $\eta$, namely $\varphi_{2,B} \sim
(\del_x + \tilde{U}^\prime) \eta_B$.  (The zero mode $u_+ \sim
\varphi_{\mathrm sol}^\prime$ does not lead to a corresponding solution
for $u_-$ and $\varphi_2$
since $(\del_x  + \tilde{U}^\prime) \varphi_{\mathrm sol}^\prime = 0)$.
The discrete spectrum of the fermions in the N=(2,2) model is then
as follows:  in the sector with $Re\, \j_+$ and $Re\, \j_-$ there
are one discrete state with zero energy $(Re\, \j_+ \sim
\varphi_{\mathrm sol}^\prime)$ and bound states with energy $\omega_B$
$(Re\, \j_+ \sim  \eta_B)$. In the sector with $Re\, \j_-$ and $Im\, \j_-$ the
same normalizable solutions are found.  Thus, as expected, the massive
bosonic spectrum of small oscillations around the soliton
background is equal to the corresponding fermionic
spectrum, and consists of massive quartets. There are also one bosonic 
(for translations) and two fermionic zero modes (zero energy solutions 
of the linearized field equations). The latter are proportional to the
nonvanishing susy variations 
$\delta \j\sim\varphi_{\mathrm sol}^\prime \e$, which are 
due to the susy parameters $Re\, \e_+$ and $Im\, \e_-$.%
\footnote{One can directly determine these fermionic zero modes  
 from (\ref{lfe}) by looking for 
time-independent normalizable solutions. One finds 
then $Re\,\j_+ \sim Im\,\j_- \sim 
\exp[-\int_0^x U'(\phi_{\mathrm sol} (x')/\sqrt2) dx']$. 
These functions are indeed proportional to $\varphi_{\mathrm sol}^\prime$. }
The zero modes do not form a susy multiplet (there are two fermionic
and only one bosonic zero mode) but this poses no problem as quantization
of collective coordinates tells us that the translational zero 
mode does not correspond to a physical particle. 
The fermionic zero modes are due to translations in superspace,
namely when the susy generators $Q_\pm$ act on the superfield
$\Phi(x,\theta)=\varphi_{\mathrm sol}$.

The topological boundary conditions for the action (\ref{n2action})
are
\begin{eqnarray}
\phi(-L/2)=(-1)^p\phi(L/2); \quad
\phi^\prime(-L/2)=(-1)^p\phi^\prime(L/2); \nonumber\\
\psi(-L/2)=(-1)^q(\gamma_3)^p\psi(L/2).
\end{eqnarray}
For the continuous spectrum with $p=1$ and $q=0$ or $q=1$
boundary conditions, the quantization conditions are
\eqa
\eta:&& kL+\d(k)=2n\pi+\pi \nonumber\\
\varphi_2:&&kL+\d(k)+\theta(k)=2n\pi+\pi \nonumber\\
Re\,\psi_+,Im\,\psi_+,Re\,\psi_-,Im\,\psi_-:&&
kL+\d(k)+{1\over2}\theta(k)=2n\pi+q\pi
\eqae
\goodbreak
The one-loop corrections to the soliton mass, differentiated w.r.t. $m$,
are then given by
\eqa
\partial_m M^{(1)}&=&\partial_m \d M^{(1)}\nonumber\\
&&\hspace{-1.5cm}+{1\over2m}\int_{-\infty}^\infty
\sqrt{k^2+m^2}\left(\d'(k)+\{\d'(k)+\theta'(k)\}
-2\{\d'(k)+{1\over2}\theta'(k)\}\right){dk\over2\pi}
\eqae
where the massive bound states do not contribute because they come
in susy multiplets. The continuum states do not contribute either, because
all phase shifts clearly cancel. The counterterm $\d M$ vanishes in the
$N=2$ model since the $\eta$ and $\phi_2$ tadpole give 
$-3\over2$ and $-1\over2$
times a fermionic tadpole, respectively. 
Note that all these cancelations are in fact
independent of any particular regularization scheme since all
integrands cancel.

To decide whether the Bogomolnyi bound is saturated we now turn to the
central charges.
The super Poincar\'e charges are obtained by the Noether method and
read
 \eqa
Q_+ &=& \int (U \j_- +\j_+^* \dot{\varphi}+\j_-^* \varphi ') dx  
 \nonumber\\
Q_- &=& -\int(U \j_+ - \j_-^* \dot{\varphi} -\j_+^* \varphi ') dx 
\eqae
With topological boundary conditions the combination $Q_+ - Q_-$ is conserved,
but not $Q_+ + Q_-$,
\eq
\dot Q_+ + \dot Q_- = -2\left[U(\psi_+-\psi_-)+
(\psi_+^* + \psi_-^*)(\dot\varphi+\varphi')\right]_{L/2}
\eqe

There are no ordering ambiguities in these operators, and
using equal-time canonical commutation relations one finds the following
algebra for $A_\pm = Q_+ \pm (Q_+)^*$ and $B_\pm = Q_- \pm Q_-^*$
\eqa
\{ A_\pm , A_\pm\} &=& \pm 2 H - (Z + Z^*)  \; ; \;
\{ A_+ , A_- \} = - (Z - Z^*) \nonumber\\
\{ B_\pm , B_\pm \} &=& \pm  2 H + 2 Re\, Z \; ; \;\;\;\;
\{ B_+ , B_- \} = Z- Z^* \nonumber\\
\{ A_\pm , B_\pm \} &=& \pm 2 P ; \{ A_+, B_- \} = \{A_- , B_+ \} = 0
\eqae
The generators which are produced on the right-hand side are
 \eqa
H &=& \int^\infty_{-\infty} \left( \dot{\varphi} \dot{\varphi}^* +
\varphi^\prime \varphi^{*\prime} + U U^*  + i\j_+^* \dot{\j}_+ + i\j_-^*
\dot{\j}_- \right) d x \nonumber\\
P &=& \int _{-\infty}^{+\infty} \left( \varphi '\dot{\varphi}^* +\dot{\varphi}
\varphi^{*\prime} +i\j^*_+\j_+ '+i\j_-^*\j_- ' \right) d x \nonumber\\
Z &=& 2\int^\infty_{-\infty} U \varphi^\prime d x 
\eqae
Since an $N=1$ massless multiplet in $D=(3,1)$ 
(which is always without central
charge) becomes a massive $N=2$
multiplet in $D=(1,1)$  whose (mass)$^2$ is
equal to the square of the central charges, while the N=(1,1) susy
algebra in $D=(3,1)$ becomes an N=(2,2) susy algebra in $D=(1,1)$ with two
central charges (the generators $P_2$ and $P_3$), it is clear why massive
multiplets of $D = (1,1)$ $N=(2,2)$ models with maximal central charge are
shortened.

Since $A_+$ and $i A_-$, and 
$B_+$ and $i B_-$ are hermitian, the BPS bound is $H \geq |Re\, Z|$.  For
the soliton one has classically that $Z$ is real and $H= Re\, Z$, 
i.e., the bound is
saturated.  At the quantum level, the 1-loop corrections to $Re\, Z$ are given
by expanding $\varphi=\varphi_{\mathrm sol}/\sqrt2+\chi$ and taking
the vacuum expectation values in the soliton vacuum
 \eqa
Z &=& Z_{cl} + 
Re \int \left\{ U^\prime (\varphi_{\mathrm sol}/\sqrt2) 
\langle \chi \chi^\prime \rangle +
\half U^{\prime\prime} (\varphi_{\mathrm sol}/\sqrt2) \langle \chi \chi \rangle 
\varphi_{\mathrm sol}^\prime/\sqrt2 \right\}
dx \nonumber\\
&=& Z_{cl} + Re \int \del_x \left[ \half U^\prime 
(\varphi_{\mathrm sol}/\sqrt2 ) \langle \chi\chi \rangle
\right] dx \nonumber\\
&=& Z_{cl} + 
{m\over 2} Re\left( \langle \chi (+ \infty) \chi (+\infty)\rangle - \langle \chi (-\infty)
\chi (- \infty) \rangle \right)
\eqae
Since {\sf asymptotically}
$\langle \chi \chi \rangle = 0$ (only $\langle \chi^* \chi\rangle$ is nonzero), 
there is no correction
to the central charges.  Because there is also no correction to the mass of
the soliton, the BPS bound remains saturated.

\section{Higher loops}
\label{sec:2loop}

In this section we repeat the two-loop calculation of the mass of the soliton 
in the sine-Gordon theory \cite{verw,vega}, paying close attention this
time to possible ambiguities.  We begin with a review of the method of 
quantization of collective coordinates, focusing on possible ordering
ambiguities. For early work on quantization of collective coordinates
see \cite{ger2,ger3,ger4,jev,chrlee,gj,jac,dhn2,tomb,ger1,poly}.
For an introduction see \cite{raj}.
 
To compute the higher loop corrections to the mass of a soliton, one may use 
standard quantum mechanical perturbation theory. One expands the 
renormalized Hamiltonian 
into a free part and an interaction part, $H=H^{0}+H_{\rm int}$, and the latter is
 expanded in terms of the dimensionless
interaction parameter $\sqrt{\hbar c\lambda /m^{_2}}$
as $H_{\rm int}=\sum_{n=1}^{\infty}H_{\rm int}^{(n)}$.\footnote{Terms with
$n$ quantum fields $\eta$ contain in addition a dimensionless factor
$(\omega L/c)^{-n/2}$. We set $c=1$ but keep $\hbar$ when useful.}
 This expansion is performed both in the soliton sector and in the trivial 
sector, and the corrections to the mass of the soliton can then be evaluated
to any given order in $\hbar\lambda /m^2$ by subtracting the energy of the 
vacuum in the trivial sector from the energy of the vacuum in the soliton 
sector. For the 
two-loop corrections (themselves of order $\hbar^{2} \lambda /m$ since the 
classical energy $M_{cl}$
of the soliton is proportional to $m^{3}/\lambda$) this means that we must 
evaluate 
\begin{eqnarray}\label{71}
M^{(2)}&=&\langle {\rm sol}|H_{\rm int,sol}^{(2)}|{\rm sol}\rangle-\langle0|H_{\rm int,triv}^{(2)}|0\rangle\nonumber\\&&+
{\sum_{p}}^\prime
\frac{\langle {\rm sol}|H_{\rm int,sol}^{(1)}|p\rangle\langle p|H_{\rm
int,sol}^{(1)}|{\rm sol}\rangle}{E_{\rm sol}-E_{p}}
\nonumber\\&&
-{\sum_{p}}^\prime\frac{\langle0|H_{\rm int,triv}^{(1)}|p\rangle\langle p|H_{\rm int,triv}^{(1)}|0\rangle}{E_{0}-
E_{p}} \label{M2}
\end{eqnarray}
Here $|{\rm sol}\rangle$ is the ground state in the soliton sector
(the soliton vacuum) with classical energy $E_{\rm sol}=M_{cl}$,
$|0\rangle$ is the ground state in the trivial (nontopological) sector
with vanishing classical energy, $|p\rangle$ are the complete sets of
eigenstates of $H^{(0)}_{\rm sol}$ and $H^{(0)}_{\rm triv}$ with positive
energies $E_{p}$, and the sums extend over all excitations but do not
include the ground state. Hence $E_{\rm sol}-E_{p}$ and $E_{0}-E_{p}$
never vanish.  The fields for the quantum fluctuations in the trivial
and the topological sectors are expanded into modes with creation and
annihilation operators, and both $|0\rangle$ and $|{\rm sol}\rangle$
are annihilated by the annihilation operators.

\subsection{The quantum Hamiltonian}

To apply this approach to the soliton in sine-Gordon theory, we begin by 
defining the sine-Gordon action\footnote{To compare with \cite{vega,verw}
we use their action. It is related to the action in section~5 by
the shift $\sqrt\l\varphi_1/m_0\to\sqrt\l\varphi_1/m_0+\pi$. Note that
in \cite{verw} mass is renormalized by 
$m_{0}^{2}=m^{2}-\delta m^{2}$.}
\begin{equation}
{\cal L}=\frac{1}{2} \dot{\phi}^{2}-\frac{1}{2}(\phi^\prime)^{2}-\frac{m_{0}^{4}}{
\lambda}\left(1-\cos\frac{\sqrt{\lambda}}{m_{0}}\phi\right)
\label{sG}
\end{equation}
The action in the trivial sector is obtained by expanding $\phi$ about the 
trivial vacuum. Since for sine-Gordon theory the latter is given by $\Phi$=0, 
we obtain $\phi= \Phi +\eta=\eta$, and
\begin{equation}
{\cal L}=\frac{1}{2}\dot{\eta}^{2}-\frac{1}{2}(\eta ')^{2}-\frac{1}{2}m_{0}^{2}
\eta^{2} +\frac{1}{4!}\lambda\eta^{4}+\ldots
\end{equation}
In 1+1 dimensional linear sigma models only mass renormalization is needed, 
$m_{0}^{2}=m^{2}+\delta m^{2}$, and at the one-loop level
$\delta m^{2}$ is fixed by requiring that
the graph with a seagull loop and two external $\eta $ fields 
 cancels the contribution from $-(1/2) \delta m^{2}\eta ^{2}$. This yields
\begin{equation}
\delta m^{2}=\frac{\lambda}{2}\sum \frac{\hbar}{2\omega_{\rm vac}L}=
 \frac{\hbar \lambda }{4\pi}\int _{0}^{\Lambda}\frac{dk}{\sqrt{
k^{2}+m^{2}}}
\end{equation}
The mass $m$ is thus the physical mass of the meson
at the pole of the propagator to this order.

A complete counterterm which removes all equal-time contractions is
\cite{col}
\begin{equation}\label{75}
\Delta H=(e^{\delta m^{2}/m^{2}}-1)\frac{m^{4}}{\lambda}\int_{-\infty}^{
+\infty}\left(1-\cos\frac{\sqrt{\lambda}}{m}\phi (x)\right)dx
\end{equation}
The corrections to the physical mass at higher loop orders are then
finite, and by expanding the final result for the soliton mass in
terms of the physical mass of the mesons, 
any ambiguity due to defining a $\Delta H$
which differs from (\ref{75}) by finite terms will be eliminated.  In
particular, the contributions from other renormalization conditions
for $m$, and finite renormalization of $\lambda$ and $\eta$, should
cancel.

The Hamiltonian in the trivial sector is simply
\begin{eqnarray}
H_{\rm triv}^{(0)}&=&\int_{-\infty}^{+\infty}\left[\frac{1}{2}\Pi_{0}^{2}(x)+\frac{1}{2}
(\eta^\prime(x))^{2} +\frac{1}{2} m^{2}\eta(x)^{2}\right]dx\\
H_{\rm int,triv}&=&-\frac{1}{4!}\lambda\eta^{4}+\ldots +\frac{1}{2}\delta m^{2}\eta
^{2}+\ldots
\end{eqnarray}
where $\Pi_{0}(x)$ is the momentum canonically conjugate to $\eta (x)$.
So $H_{\rm int,triv}^{(n)}$ contains only terms for even $n$ and to obtain the 
contributions of the trivial sector to the two-loop corrections to the mass of 
the soliton we must evaluate $-\frac{1}{4!}\lambda\langle0|\eta^{4}|0\rangle+\frac{1}{2}
\delta m^{2}\langle0|\eta^{2}|0\rangle$. Note that there are no ordering ambiguities in 
the Hamiltonian of the trivial sector. 

To obtain the Hamiltonian in the soliton sector, one must use the formalism 
for quantization of collective coordinates \cite{raj}. 
Although the final formulas look somewhat
complicated, the basic idea is very simple: one expands $\phi (x,t)$ again 
into a sum of a background field (the soliton) and a complete set of small
fluctuations about the background field, but instead of simply writing
$\phi (x,t)=\phi_{\rm sol}(x)+\sum q^{m}(t)\eta_{m}(x)$ where $\eta_{m}(x)$ stands 
for all modes (eigenfunctions of the linearized field equations), 
one deletes the zero mode for translations from the sum, and 
reintroduces it by replacing $x$ by $x- X(t)$ on the right hand side of the 
expansion of $\phi$. For small $X(t)$, the expansion of $\phi_{\rm sol}(x-X(t))$ 
into a Taylor series gives $\phi_{\rm sol}-X(t)\phi'_{\rm sol}(x,t)+...$, 
and since $\phi '_{\rm sol}
(x,t)$ is  the translational zero mode (the solution of the linearized field 
equations with vanishing energy), one has not lost any degrees of freedom.
Hence one substitutes
\begin{equation}
\phi (x,t)= \phi_{\rm sol}(x-X(t))+{\sum}^\prime q^{m}(t)\eta_{m}(x-X(t))
\end{equation}
into the action in (\ref{sG}), and using the chain rule, one finds an action 
of the form of a quantum mechanical nonlinear sigma model (but with
infinitely many degrees of freedom)
\begin{equation}
L=\dot{u}^{I}g_{IJ}(u)\dot{u}^{J}-V(u);u^{I}=\{ X(t),q^{m}(t)\}
\end{equation}
The metric $g_{IJ}$ is given by
\eq
g_{IJ}=\int{\partial\phi(x,t)\over\partial u^I}
{\partial\phi(x,t)\over\partial u^J}dx
\eqe
and contains space integrals over expressions which depend on 
$q^{m}(t), \eta_{m}(x)$ and $\phi_{\rm sol}(x)$, but not on $X(t)$ due to the 
translational invariance of the integral over $x$. The Hamiltonian is then
 simply given by 
\begin{equation}
\label{qmham}
H=\pi_{I}g^{IJ}(u)\pi_{J}+V(u);\pi_{I}=\{ P(t),\pi_{m}(t)\}
\end{equation}
where $g^{IJ}(u)$ is the matrix inverse of the metric $g_{IJ}(u)$ and
$P(t)$ is the center of mass momentum (the momentum conjugate to
$X(t)$),while $\pi_{m}(t)$ are momenta canonically conjugate to
$q^{m}(t)$.

Classically, this is the whole 
result. One may check that the equal-time Poisson brackets $\{ Q,P \}=1,\{ 
q^{m},\pi_{n} \} =\delta_{n}^{m}$ imply $\{\phi(x),\Pi_{0}(y)\}=\delta (x-y)$
where $\Pi_{0}(x,t)=\dot{\phi}(x,t)$, and vice-versa. Hence, the 
transition from
$\phi(x,t)$ and $\Pi_{0}(x,t)$ to $\{ X(t),q^{m}(t)\}$ and $\{ P(t),\pi_{m}(t)
\}$ is a canonical transformation. It is useful to recast the ``quantum 
mechanical'' Hamiltonian in (\ref{qmham}) into a form which resembles more 
 the Hamiltonian of a 1+1 dimensional field theory.
To this purpose we introduce fields constructed from $q^{m}$ and $\pi_{m}$
as follows
\begin{eqnarray}
\eta(x,t)&\equiv& \sum\,\! ' q^{m}(t)\eta_{m}(x-X(t))\\
\pi(x,t)&\equiv&\sum\,\! ' \pi_{m}(t)\eta_{m}(x-X(t))
\end{eqnarray}
By combining the $\pi_{m}$ and $q^{m}$ with the functions $\eta_{m}(x)$ which 
appear in $g^{IJ}(u)$, one can write the complete Hamiltonian only in terms 
of the fields $\eta(x,t)$ and $\pi(x,t)$ and the background field 
$\phi_{\rm sol}(x)$. To simplify the notation, we introduce an inner product
$(f,h)\equiv\int_{-\infty}^{+\infty}f^{*}(x)h(x)dx$. Note that the functions 
$\eta_{m}$ which parameterize the small fluctuations are orthogonal to the
zero mode  $\phi_{\rm sol}'$ since they correspond to different
eigenvalues of the kinetic operator 
\begin{equation}
(\phi_{\rm sol}',\eta_{m})=0
\end{equation}
a result we shall use repeatedly. The classical Hamiltonian density ${\cal
H}=T_{00}$ is given by
\begin{eqnarray}
{\cal H}&=&\frac{1}{2} \Pi_{0}^{2}(x,t)+\frac{1}{2}\phi '(x,t)^{2}+V(\phi)
\nonumber \\&=&\frac{1}{2}\pi^{2}(x-X(t),t)-\pi(x-X(t),t)\frac{ P+(\pi,\eta '
) }{M_{cl}[1+(\eta ',\phi_{\rm sol}')/M_{cl}]}\phi_{\rm sol}'(x-X(t))\nonumber \\
&+&\frac{[P+(\pi,\eta ')]^{2}}{2 M_{cl}^{2}(1+(\eta ',\phi_{\rm sol}')/M_{cl})}
[\phi_{\rm sol}'(x-X(t))]^{2}+\nonumber \\
&+&\frac{1}{2}[\eta '(x-X(t),t)+\phi_{\rm sol}'(x-X(t))]^{2}+V(\phi_{\rm sol}+\eta)
\end{eqnarray}
A great simplification occurs in 
the Hamiltonian $H=\int_{-\infty}^{+\infty}{\cal H}dx$ because due to the
orthogonality of the zero mode $\phi_{\rm sol}'$ to the fluctuations $\eta_m$, the 
complicated second term in ${\cal H}$ cancels.
There should be no terms linear in the fluctuations $\eta$ and $\pi$
and the collective coordinates $X$, $P$ in $H$ (i.e., after 
integrating ${\cal H}$ over $x$). That this is indeed the case follows
 from the field equation
\begin{equation}
\phi_{\rm sol}''=V'(\phi_{\rm sol})
\end{equation}
The classical energy of the soliton at rest is given by
\begin{equation}
M_{cl}=\int_{-\infty}^{+\infty}(\phi_{\rm sol}')^{2}dx
\end{equation}
which follows from equipartition of energy 
\begin{equation}
\frac{1}{2}(\phi_{\rm sol}')^{2}=V(\phi_{\rm sol})
\end{equation}

Thus we arrive at the following expression for the classical Hamiltonian
in the topological sector
\begin{eqnarray}\label{89}
H_{\rm sol}^{(0)}&=&M_{cl}+\int_{-\infty}^{+\infty}\left[\frac{1}{2}\pi(x,t)^{2}+
\frac{1}{2}\eta '(x,t)^{2}+\frac{1}{2}\eta^{2}V''(\phi_{\rm sol})\right]dx\\
\label{90}
H^{\rm cl}_{\rm int,sol}&=&\frac{1}{2M_{cl}}\frac{[P+(\pi,\eta ')]^{2}}{1+(\eta ',
\phi_{\rm sol}')/M_{cl}}+\int \left[\frac{1}{3!}\eta^{3} V'''(\phi_{\rm sol})+
\right.
\nonumber\\&& \left.
+\frac{1}{4!}\eta^{4}V''''(\phi_{\rm sol})+\ldots\right]dx
\end{eqnarray}
All $X(t)$ dependence has disappeared from $H$ due to translational invariance 
of the integration over $x$.

We must now discuss the subtle issue of operator ordering in $H$. We shall 
consider a soliton at rest, so we set $P=0$. Furthermore, due to $[q^{m}, 
\pi_{n}]=i\hbar\delta_{n}^{m}$ and $(\eta_{m},\eta_{m}')=0$ 
(since we work in a finite volume, and $\eta$ and $\eta'$ have the
same boundary conditions)
one has the
 equality $(\pi,\eta ')=(\eta ',\pi)$, at least if one considers a finite 
number of modes in $\eta$ and $\pi$. However, there are operator ordering 
ambiguities both in $(\pi, \eta ')^{2}$ and also with respect to the term 
$(\eta ', \phi_{\rm sol} ')/M_{\rm cl}$ in the denominator.

In general, one may require that the generators $H, P=\int T_{01}dx$ and $L=
\int x T_{00} dx$ satisfy the Poincar\'e algebra \cite{tomb}%
\footnote{The Noether current for the orbital part of the angular
momentum $J_{\rho\sigma}$ is given by $j^{\mu}_{\rho \sigma}=(x_{\rho} T^{\mu}_
{\sigma}-x_{\sigma}T_{\rho}^{\mu})$ so $J_{01}=L=\int (x_{0}T^{0}_{1}-x_{1}
T^{0}_{0})dx$. At t=0 this reduces 
to $\int xT_{00} dx$.}. 
The expressions for these operators are quite complex, and in general it
seems likely that the operator ordering which leads to closure of the Poincar\'e
algebra is unique (in quantum gravity, such an ordering has never been found).
There is, however, an ordering which guarantees closure, and this is the 
ordering we shall adopt. It is obtained by making the canonical transformation 
at the quantum level. One begins with the quantum Hamiltonian in 
``Cartesian coordinates'' (i.e., in terms of the operators $\Pi_{0}(x)$ and 
$\phi (x)$). In the Schr\"{o}dinger representation the operator
$\Pi_{0}(x)$ is represented by $\partial/\partial
\phi(x)$, and making the change of coordinates  from $\phi(x)$ to $X$ 
and $q^{m}$, one obtains  the Laplacian in curved space by applying the chain 
rule
\begin{equation}
\sum\left(\frac{\partial}{\partial \alpha^{i}}\right)^{2}=\frac{1}{\sqrt{g}}\frac
{\partial}{\partial u^{I}}\sqrt{g(u)}g^{IJ}(u)\frac{\partial}{\partial u^{J}}
\end{equation}
where $\alpha^{i}$ is the set $\phi(x)$ and $u^{I}$ the set $X$, $q^{m}$. If the
inner product in $\alpha$ space is given by $(f,h)=\int f^{*}(\alpha)h(\alpha)
(\prod d\alpha^{i})$, it becomes in $u$ space 
$(f,h)=\int f^{*}(\alpha(u))h(\alpha(
u))\sqrt{g(u)}(\prod du^{I})$. With this inner product, the relation between 
$\partial /\partial u^{J}$ and the conjugate momenta $\pi_{J}$ in the 
Schr\"{o}dinger representation is not
simply $\pi_{J}=\partial /\partial u^{J}$ , but rather 
\begin{equation}
\label{trans}
\frac{\partial}{\partial u^{J}}=g^{1/4}(u)\pi_{J} g^{-1/4}(u)
\end{equation}
as one may check.\footnote{In \cite{raj} 
this derivation of the operator ordering of the Hamiltonian
 is given, but at the end $\partial /\partial u^{J}$ is replaced by $\pi_{J}$
which is incorrect. In \cite{ger1} the correct quantum Hamiltonian is obtained, 
but
the factors $g^{-1/4}(u)$ are produced by ``Redefining the Hilbert space so 
as to eliminate the measure from this scalar product...''. We claim that
the relation (\ref{trans}) is not a  convention or a choice of basis,
but is fixed because the inner product has been specified.}
 Hence the correct quantum Hamiltonian is given by
\begin{equation}
\hat{H}=\frac{1}{2}\frac{1}{g(u)^{1/4}}\pi_{I}\sqrt{g(u)}g^{IJ}(u)\pi_{J}\frac
{1}{g(u)^{1/4}}+V(u)+\Delta H
\end{equation}
with $\Delta H$ given by (\ref{75}).

 It is often useful to {\em rewrite} this Hamiltonian such that all 
expressions are 
Weyl ordered, because then one can use Berezin's theorem and find at once the 
action to be used in the path integral. The result is
\begin{eqnarray}
\hat{H}&=&\frac{1}{2}(\pi_{I}g^{IJ}\pi_{J})_{W}+V(u)+\Delta V +\Delta H
\\ \Delta V&=& \frac{\hbar^{2}}{8}\left[\partial_{I}\partial_{J}g^{IJ}(u)-4g^{-1/4}
(u)\partial_{I}\{ g^{1/2}(u)g^{IJ}(u)\partial_{J}g^{-1/4}(u)\}\right]
\end{eqnarray}
The operator $(1/2)\left(\pi_Ig^{IJ}\pi_J\right)_W$ is obtained by
promoting (\ref{89}) and (\ref{90}) to operators and Weyl ordering.
Weyl ordering yields then $(1/2)((1/4)\pi_I\pi_Jg^{IJ}+
(1/2)\pi_Ig^{IJ}\pi_J+(1/4)g^{IJ}\pi_I\pi_J)$.
Substituting the expression for $g^{IJ}(u)$ \cite{raj} one finds
\begin{eqnarray}
\partial_{I}\partial_{J}g^{IJ}&=&\partial_{q^{m}}\partial_{q^{n}}\left\{ \frac{
(\eta^{m},\eta ')(\eta^{n}, \eta ')}{(\psi_{0}, \phi_{\rm sol}'+\eta ')^{2}}\right\}
\nonumber\\ 
4g^{-1/4}\partial_{I}\left\{ g^{1/2}g^{IJ}\partial_{J}g^{-1/4}\right\}
&=&\frac{1}{(\psi_{0},\phi ')^{1/2}}\frac{\partial}{\partial q^{m}}\left[\frac{1}{
(\psi_{0},\phi ')^{3/2}}\frac{\partial}{\partial q^{m}}(\psi_{0},\phi ')^{2}+
\right.\nonumber\\&&\left.
+\frac{(\eta ',\eta_{m})(\eta ', \eta_{n})}{(\psi_{0},\phi ')^{7/2}}\frac{
\partial}{\partial q^{n}}(\psi_{0},\phi ')^{2}\right]
\end{eqnarray}
where $\phi=\phi_{\rm sol}+\eta$, and $\psi_{0}=\phi_{\rm sol} '/\sqrt{M_{\rm cl}}$ is the 
normalized zero mode. This leads to
\begin{eqnarray}
 \Delta V&=&\frac{\hbar^{2}}{8}\left[-\frac{(\psi_{0},\eta_{m}')(\eta_{m}',\psi_{0})
}{(\psi_{0},\phi ')^{2}} \right.\nonumber\\&&
-2\frac{ (\psi_{0},\eta_{m}')(\eta_{m},\eta_{n}')(
\eta_{n}, \eta ')+(\psi_{0},\eta_{m}')(\eta_{m},\eta ')(\eta_{n},\eta_{n}') }
{(\psi_{0},\phi ')^{3}}\nonumber\\
&&\left.+\frac{\{(\psi_{0},\eta_{m} ')(\eta_{m},\eta ')\}^{2}}{(\psi_{0},\phi ')^{4}}
+\frac{(\eta_{m},\eta_{n}')(\eta_{n},\eta_{m} ')+(\eta_{m},\eta_{m} ')^{2}}
{(\psi_{0},\phi ')^{2}} \right]
\end{eqnarray}
Further simplifications result by using the identities
\begin{eqnarray}
&&(\psi_{0},\psi_{0}')=0,\;(\psi_{0},\eta ')=(\psi_{0},\phi '),\; (\psi_{0},
\eta_{m}')=-(\psi_{0}',\eta_{m})\nonumber\\&&(\eta_n,\eta')=(\eta_n,\phi'),\;
(\eta_{m},\eta_{m}')=0,\;
{\sum}^\prime \eta_{m}(x)\eta_{m}(y)=\delta(x-y)-\psi_{0}(x)\psi_{0}(y)
\end{eqnarray}
The final answer for $\Delta V$ reads then 
\begin{eqnarray}
\Delta V&=&\frac{\hbar^{2}}{8}\left[ -\frac{(\psi_{0}',\psi_{0}')}{(\psi_{0},\phi 
')^{2}} + 2\left\{\frac{(\psi_{0}',\phi '')}{(\psi_{0},\phi ')^{3}} -
\frac{(\psi_{0}
',\psi_{0}')}{(\psi_{0}',\phi ')^{2}}\right\}\right.\nonumber\\&&\left.
 + \frac{(\psi_{0}',\phi ')^{2}}
{(\psi_{0},\phi ')^{4}} -\sum_{m,n}\frac{|(\eta_{m},\eta_{n}')|^{2}}
{(\psi_{0}',\phi ')^{2}}\right]
\end{eqnarray}  

The total Hamiltonian is then the sum of $H^{(0)}_{\rm sol}$ in
(\ref{89}) in which no ordering 
problems are present, and $\left(H^{\rm cl}_{\rm
int,sol}\right)_W +\Delta V+\Delta H$ with $\left(H^{\rm cl}_{\rm
int,sol}\right)_W$ given by (\ref{90}) with the complicated momentum
dependent term Weyl-ordered.This is the result in \cite{ger1}. 
A drastic 
simplification is obtained by rewriting the latter term in a particular
non-Weyl-ordered way in such a way that it absorbs all terms in $\Delta V$ 
except the first
one \cite{tomb}. This leads to the final form of the interaction 
Hamiltonian\footnote{
The first term in (\ref{102}) can be written for $P=0$ as $[(\pi,\eta '/F)+(\eta'
/F,\pi)]^{2}$, where $F=1+(\eta ',\phi_{\rm sol}')/M_{\rm cl}$. 
Weyl-ordering of $(\pi , \eta '/F)^{2}$ yields
\[
\frac{1}{4}
\int_{-\infty}^{+\infty}\left[\pi (x)(\pi,\eta '/F) \frac{\eta '}{F} (x)+\frac
{\eta '}{F} (x)(\eta '/F,\pi)\pi (x)\right] dx +2(\pi,\eta '/F)(\eta '/F,\pi)
\]

Evaluating the difference of the two expressions one needs the following 
commutators
\begin{eqnarray}
[\pi (x),\eta '(y)] &=&-i\partial_{y}\delta (x-y)+\frac{i}{M_{\rm cl}}\phi ' _
{\rm sol}(x)\phi''_{\rm sol}(y)\nonumber \\ \;
[\pi (x),1/F] &=&-i\phi ''_{\rm sol}(x)/(M_{\rm cl}F^{2})
\nonumber
\end{eqnarray}

Straightforward algebra produces then all terms in $\Delta V$ except the
 first one.
}
\begin{eqnarray}\label{102}
H_{\rm int,sol}&=&\frac{1}{8 M_{\rm cl}}\left\{ (P+(\eta ',\pi)),\frac{1}{1+(\eta ',\phi 
'_{\rm sol})/M_{\rm cl}}\right\}^{2}\nonumber \\&&
-\frac{\hbar^{2}}{8 M_{\rm cl}^{2}}\int_{-\infty}^{+\infty}\frac{(\phi ''_{\rm sol})
^{2} dx}{[1+(\eta ',\phi_{\rm sol}')/M_{\rm cl}]^{2}}+\Delta H
\nonumber\\&&
+\int_{-\infty}^{+\infty}\left[\frac{1}{3!}\eta^{3}
V'''(\phi_{\rm sol})+\frac{1}{4!}\eta^{4}V''''(\phi_{\rm sol})+\ldots\right]dx
\end{eqnarray}
Note that the first term is the square of a Weyl-ordered operator, but
is not itself Weyl-ordered.

For the two-loop calculation we are going to perform we set P=0, and we 
only need the terms as far as quartic in $\eta $ and $\pi$. This leads to
\begin{eqnarray}
H_{\rm int,sol}&=&\frac{1}{2M_{\rm cl}}\left(\int_{-\infty}^{+\infty}\eta '\pi dx\right)
\left(\int_{-\infty}^{+\infty}\eta '\pi dy\right)
-\frac{\hbar^{2}}{8 M_{\rm cl}^{2}}\int_{-
\infty}^{+\infty}(\phi ''_{\rm sol}(x))^{2} dx \nonumber \\&&
+\Delta H
-\frac{\sqrt{\lambda}m}{3!}\int_{-\infty}^{+\infty}\eta^{3}(x)\sin
\frac{\sqrt{\lambda}}{m}\phi_{\rm sol}(x)dx \nonumber\\&&-
\frac{\lambda}{4!}
\int_{-\infty}^{+\infty}\eta^{4}(x)\cos\frac{\sqrt{\lambda}}{m}\phi_{\rm sol}(x)dx
+\ldots
\label{hint}
\end{eqnarray}
The counterterms are the same in the topological sector as in the trivial sector,
but in the trivial sector we decomposed $\phi=\Phi +\eta=\eta$, while here 
we expand $\phi=\phi_{\rm sol}+\eta $. We obtain then
\begin{eqnarray}
\Delta H&=&\frac{m^{4}}{\lambda}\left\{e^{\frac{\delta m^{2}}{m^{2}}}-1\right\}
\int_{-\infty}^{+\infty}\left\{1-\cos \frac{\sqrt{\lambda}}{m}\phi (x)\right\}dx 
\;\nonumber\\ \;
&=&\frac{m^{4}}{\lambda}\left[\frac{\delta m^{2}}{m^{2}}+\frac{1}{2}\left(\frac{\delta
 m^{2}}{m^{2}}\right)^{2}\right]\int_{-\infty}^{+\infty}\left\{ 1-\cos\frac{\sqrt{\lambda}}{m}
\phi_{\rm sol}(x)\right\}   dx \nonumber\\
\;&+&\frac{\delta m^{2}m}{\sqrt{\lambda}}\int_{-\infty}^{+\infty}\eta (x)
\sin\frac{\sqrt{\lambda}}{m}\phi_{\rm sol}(x)dx\nonumber\\&&
+\frac{1}{2}\delta m^{2}\int_{-\infty}^{+\infty}\eta ^{2}(x)\cos\frac
{\sqrt{\lambda}}{m}\phi_{\rm sol}(x)dx+...
\label{hct}
\end{eqnarray}
The first term is the counterterm for the one-loop graphs and will not 
contribute to our two-loop calculation.  

\subsection{The actual two-loop calculation}

Using the explicit expressions for the classical soliton solution
\begin{equation}
\phi_{\rm sol}=\frac{4m}{\sqrt{\lambda}} \arctan (e^{mx})
\end{equation}
one finds $M_{\rm cl}=\frac{8m^{3}}{\lambda}$ and
\begin{eqnarray}
\sin \left(\frac{\sqrt{\lambda}}{m}\phi_{\rm sol}\right)&=&-4\frac{e^{mx}-e^{-mx}}{(e^{mx}
+e^{-mx})^{2}};\;\cos \frac{\sqrt{\lambda}}{m}\phi_{\rm sol}=1-\frac{8}{(e^{mx}+
e^{-mx})^{2}}\nonumber\\&&
-\frac{1}{8M_{\rm cl}^{2}}\int_{-\infty}^{+\infty}(\phi_{\rm sol}''(x))^{2}dx=-\frac
{\lambda}{192 m}
\end{eqnarray}
Substituting these results into (\ref{hint},\ref{hct}), 
we find for the Hamiltonian
\begin{eqnarray}
H_{\rm int,sol}^{(1),I}&=&-\frac{\sqrt{\lambda}m}{6}\int_{-\infty}^{+\infty}\frac{
(-4)(e^{mx}-e^{-mx})}{(e^{mx}+e^{-mx})^{2}}\eta^{3}(x) dx \label{H1I}\\
H_{\rm int,sol}^{(1),II}&=&\frac{\delta m^{2} m}{\sqrt{\lambda}}\int_{-\infty}^{
+\infty}\frac{(-4)(e^{mx}-e^{-mx})}{(e^{mx}+e^{-mx})^{2}}\eta (x)dx\\
H_{\rm int,sol}^{(2),I} &=&-\frac{\lambda}{24}\int_{-\infty}^{+\infty}[1-
\frac{8}{(e^{mx}+e^{-mx})^{2}}]\eta^{4}(x)dx\\
H_{\rm int,sol}^{(2),II}&=&\frac{1}{2}\delta m^{2} \int_{-\infty}^{+\infty}[1-
\frac{8}{(e^{mx}+e^{-mx})^{2}}]\eta^{2}(x) dx\\
H_{\rm int,sol}^{(2),III}&=&\frac{\lambda}{16 m^{3}}(\int_{-\infty}^{+\infty}
\eta '(x)\pi (x)dx)^{2}\\
H_{\rm int,sol}^{(2),{\rm rest}}&=&-\frac{\lambda \hbar^{2}}{192 m}+\frac{(\delta m^{2})
^{2}}{2\lambda}\frac{4}{m}
\end{eqnarray}

We now put the system in a box of length $L$.
We expand $\eta (x)$ into creation and annihilation operators
\begin{eqnarray}
\eta(x)&=& \sum_{n=-\infty}^{+\infty}\sqrt{\frac{\hbar}{2\omega_{n}}}
(a(q_{n})E(q_{n},x)e^{i k_{n} x}+h.c.)\\
E(q_{n},x)&=&\frac{i\tanh mx +q_{n}}{[(1+q_{n}^{2})L-\frac{2}{m}\tanh \frac{
mL}{2}]^{1/2}}\\
q_{n}&=&\frac{k_{n}}{m};\omega_{n}^{2}=k_{n}^{2}+m^{2}
\end{eqnarray}
The functions $E(q_{n},x)e^{ik_{n}x-i\omega_{n}t}\equiv \eta_{n}(x)e^{-i
\omega_{n}t}$ are eigenfunctions of $H^{(0)}_{\rm sol}$ and
satisfy the linearized field equations
\begin{equation}
[\partial_{x}^{2}+\omega_{n}^{2}-m^{2}(1-2\cosh ^{-2}(x))]\eta_{n}=0
\end{equation}
and for large $q_{n}$ they tend to $\frac{1}{\sqrt{L}}$ which leads to the
familiar normalization factor $(2\omega_{n}L)^{-1/2}$ for free fields.
At $x=\pm L/2$ we find the phase shift
\begin{equation}
\delta (k)=2 \arctan\{ \frac{m}{k}\tanh 
\frac{1}{2}mL\}\label{deltaka} 
\end{equation}
The momenta $k_{n}$ are discretized by adopting antiperiodic 
boundary conditions\footnote{Of course, we differ here from refs.
\cite{vega,verw} who choose periodic boundary conditions. The
necessity for the antiperiodic boundary conditions has been
discussed at length in the preceding sections.}
\eq
k_{n}L+\delta(k_{n})=2\pi n+\pi
\label{pb}
\eqe
As eigenfunctions of a self-adjoint operator on a compact space,
the functions $E(q_{n},x)$ have the same orthogonality properties as
plane waves and they are normalized to unity,
for example $\int_{-L/2}^{+L/2}|E(q_{n},x)|^{2}dx=1$.
As in the case of the kink $\delta(k)$ in (\ref{deltaka}) is
discontinuous. The values of $n=\ldots,-2,-1,0,1,2\ldots$
give solutions of (\ref{pb}) $kL=
\ldots,-2\pi,0,0,2\pi,4\pi,\ldots$. Again, we see a defect in
the $n$ to $k$ mapping: the $k=0$ solution is obtained
for $n=-1$ and $n=0$. 
If we map $k=0$ onto $n=-1$, the remaining number $n=0$ can be assigned
to the only discrete solution (the translational zero mode). Then using
the Euler-Maclaurin formula the
sum over $n$ can be converted into integral over $k$ with continuous
measure. We now 
evaluate the various terms in (\ref{M2}), adding subsets of terms
which combine to cancel divergences.

\subsubsection*{The $\eta^{4}$, $\delta m^{2}\eta^{2}$, and $(\delta m^2)^2$ terms in $H_{\rm int,sol}^
{(2)}$ and $H_{\rm int,vac}^{(2)}$}
Since there are 3 ways to contract the 4 $\eta$'s, one finds
from the term with $\eta^4$
\begin{eqnarray}  
&&\langle {\rm sol}|H_{\rm int,sol}^{(2),I}|{\rm sol}\rangle=-\frac{\lambda}{8}\int_{-\infty}^{+\infty}\left[1-
\frac{8}{(e^{mx}+e^{-mx})^{2}}\right]
\sum_{n,p=-\infty}^{+\infty\prime}\nonumber
\\&& \frac{(\tanh^{2}mx-1)+(q_{n}^{2}+1)}{(1+q_{n}^{2})L-\frac{2}{m}\tanh\frac
{mL}{2}}\frac{(\tanh^{2}mx-1)+(q_{p}^{2}+1)}{(1+q_{p}^{2})L-\frac{2}{m}\tanh 
\frac{mL}{2}}\frac{dx}{2\omega_{n}2\omega_{p}}
\end{eqnarray}
Using the integrals in the appendix, we record the contributions due to 
terms with none, one and two factors $\tanh^{2} x-1$ in 3 separate lines
\begin{eqnarray}
&=&-\frac{\lambda}{32}[ \frac{1}{L}(\sum\frac{1}{\omega})^{2}-\frac{4}{mL^{2}}
(\sum\frac{1}{\omega})^{2}+\{ \frac{4}{L^{2}}(\sum\frac{1}{\omega})
(\sum\frac{1}{\omega^{2}})+...\}\nonumber\\&&
+\frac{4}{3}\frac{m}{L^{2}}(\sum\frac{1}{\omega})(\sum\frac{1}{\omega^{3}})+
\{\frac{8}{3}\frac{1}{L^{3}}(\sum\frac{1}{\omega^{3}})^{2}+\frac{8}{3}\frac{1}
{L^{3}}(\sum\frac{1}{\omega})(\sum\frac{1}{\omega^{5}})+...\}\nonumber\\
&&-\frac{4m^{3}}{5L^{2}}(\sum\frac{1}{\omega^{3}})^{2}+\{-\frac{16 m^{2}}{5
L^{3}}(\sum\frac{1}{\omega^{3}}(\sum\frac{1}{\omega^{5}})+...\} ]
\label{res}
\end{eqnarray}

The first two terms come from the factor $1-8/(e^{mx}+e^{-mx})^{2}$ whose 
$x$-integral is equal to $L-\frac{4}{m}$. Clearly, there is a divergence 
proportional to $L$ which will be cancelled by the corresponding contribution 
from the vacuum sector. The integrals of $1-8/(e^{mx}+e^{-mx})^{2}$ times
$(\tanh^{2} x-1)^{p}$ for $p=1$ and $p=2$ are given by $\frac{2}{3m}$ and
$-\frac{4}{5m}$, respectively, and do not contain divergences which are due 
to the $x$ integral, but they still contain divergences due to the
sums over $n$ and $p$. The terms inside curly brackets  are due to expanding the denominators 
$(1+q_{n}^{2})L-\frac{2}{m}\tanh \frac{mL}{2}
\sim(1+q_n^2)L-\frac2m$. (We have already set 
$\tanh \frac{mL}{2}=1$ since the difference vanishes exponentially fast for $
L\rightarrow\infty$). Not all these terms vanish for $L\rightarrow \infty$,
but they will cancel with similar terms from other contributions. The sums 
\begin{equation}
\sum \frac{1}{\omega}\equiv\sum_{-\infty}^{+\infty}\,\! '\frac{1}{\omega_{n}}
\sim \frac{L}{2\pi}2\int_{0}^{\infty}\frac{1}{\omega (k)}\left[1+\frac{1}{L}\delta '
(k)\right]dk
\end{equation}
will get contributions from $\delta '(k)$, but by first combining such sums,
we shall find that only differences like $\sum\frac{1}{\omega}-\sum\frac{1}{
\tilde{\omega}}$ occur, and this will simplify the analysis significantly. We 
wrote an approximation symbol $\sim$ in $(\sum\frac{1}{\omega})$ because the 
evaluation of such sums has been found to be regularization dependent 
at one-loop level. This is the issue we want to study now at the two-loop level.

The evaluation of $\langle {\rm sol}|H_{\rm int,sol}^{(2),II}|{\rm sol}\rangle$ is straightforward and
yields
\begin{eqnarray}
&&\frac{1}{2}\delta m^{2}\int_{-\infty}^{+\infty}[1-\frac{8}{(e^{mx}+e^{-mx})
^{2}}]\sum_{n}\frac{(\tanh^{2}mx-1)+(q_{n}^{2}+1)}{(q_{n}^{2}+1)L-\frac{2}{m}}
\frac{dx}{2\omega_{n}}\nonumber\\
&=&\frac{1}{2}\delta m^{2}(L-\frac{4}{m})(\frac{1}{2L}\sum\frac{1}{\omega}
+\{ \frac{1}{mL^{2}}\sum\frac{1}{\omega^{3}}+...\} )\nonumber\\
&&+\frac{1}{2}\delta m^{2}(\frac{2}{3m})\frac{m^{2}}2\frac1L\sum{1\over\omega^3}
\label{com}
\end{eqnarray}
where the last line contains the contributions from the terms with $\tanh^{2}x
-1$. 

The contributions from the trivial sector are only $\frac{1}{2} \delta m^{2}
\langle0|\eta^{2}|0\rangle-\frac{\lambda}{4!}\; \langle0|\eta^{4}|0\rangle$, where $\eta (x)$ are now
simple plane waves and they yield
\begin{equation}
\langle0|H_{\rm triv}+\Delta H_{\rm triv}|0\rangle=-\frac{1}{2}\delta m^{2}\sum\frac{1}{2\tilde{
\omega}}-\frac{\lambda}{8}(\sum\frac{1}{2\tilde{\omega}})^{2}\frac{1}{L}
\end{equation}

Finally there is the contribution from the term proportional to $(\delta m^{2}
)^{2}$ in $\Delta H_{\rm sol}$; it yields
\begin{equation}
\langle {\rm sol}|\frac{1}{2}\frac{(\delta m^{2})^{2}}{\lambda}\int_{-\infty}^{+\infty}
[1-\cos \frac{\sqrt{\lambda}}{m}\phi_{\rm sol}(x)]dx=\frac{\lambda}{8m L^{2}}
(\sum\frac{1}{\tilde{\omega}})^{2}
\end{equation}
where we wrote $\delta m^{2}$ as $\frac{\lambda}{4L}\sum\frac{1}{\tilde{\omega}
}$. 

We first demonstrate that the terms due to expanding the denominators 
$(1+q_{n}^{2})L-\frac{2}{m}$ cancel. Consider as an example  the last term 
in the first line of (\ref{res}). It cancels the last term in the first line 
of (\ref{com}).To see this, write $\delta m^{2}$ into its original form as
a sum over modes
\begin{equation}
\delta m^{2}=\frac{\hbar\lambda}{4L}(\sum\frac{1}{\tilde{\omega}})
\end{equation}
One finds for the terms with $L^{-2}$
\begin{eqnarray}
&&(-\frac{\lambda}{32})\frac{4}{L^{2}}(\sum\frac{1}{\omega})(\sum\frac{1}
{\omega^{3}}-\frac{1}{2}(-\frac{\lambda}{4L}\sum\frac{1}{\tilde{\omega}})\frac
{1}{mL}\sum\frac{1}{\omega^{3}}\nonumber\\
&&=-\frac{\lambda}{8L^{2}}(\sum\frac{1}{\omega}-\sum\frac{1}{\tilde{\omega}})(
\sum\frac{1}{\omega ^{3}})
\end{eqnarray}
Note now that $\frac{1}{L}\sum\frac{1}{\omega^{3}}$ is finite, while
\begin{eqnarray} 
\frac{1}{L}(\sum\frac{1}{\omega}-\sum\frac{1}{\tilde{\omega}})&=&\frac{1}
{L}[(\sum_{-N}^{-1}+\sum_{1}^{N})\frac{1}{\omega_{n}}-\sum_{-N}^{N}\frac{1}
{\tilde{\omega}_{n}}]\nonumber\\&=&\frac{1}{2\pi}\int_{-\infty}^{+\infty}\frac{
1}{\sqrt{k^{2}+m^{2}}}(\frac{\delta (k)}{L})dk\rightarrow 0
\label{dif}
\end{eqnarray}
is unambiguous and vanishes (the absence of the $n=0$ contribution
allows the application of the Euler-Maclaurin formula as discussed after
(\ref{pb})). Hence, the contributions due to 
expanding the denominator $(q_{n}^{2}+1)L-\frac{2}{m}$ cancel. Similar 
cancelations  occur in other pairs of corresponding terms, and we shall
therefore only use the terms $(q_{n}^{2}+1)L$ in the denominators of 
$E(x,q_{n})$ from now on.

The sum of all contributions from the terms with $\eta^{4}$, $\delta
m^{2}\eta^{2}$, and $(\delta m^2)^2$
in the topological and trivial sectors is then found to combine 
into differences, except for one term, namely the contribution from the first
term in the third line in (\ref{res}) 
\begin{eqnarray}
\langle H^{(2)}\rangle \Big|_{\eta^{4}\; {\rm and} \;\delta m^{2}\eta^{2}}
=&& \nonumber\\
\epsfysize 20pt\epsfbox{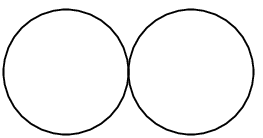}
&\raisebox{8pt}{$+$}&
\epsfysize 20pt\epsfbox{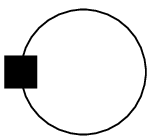}
\hspace{4em} \raisebox{8pt}{$-$}\quad
\raisebox{8pt}{\Big\{}\quad
\epsfysize 20pt\epsfbox{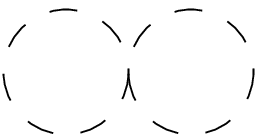}
\quad\raisebox{8pt}{$+$}\quad
\epsfysize 20pt\epsfbox{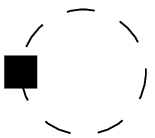}
\quad\raisebox{8pt}{\Big\}}
\quad\raisebox{8pt}{$+\quad
(\epsfysize 6pt\epsfbox{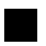})^2=$}
\nonumber\\
-\frac{\lambda}{32 L} (\sum\frac{1}{\omega})^{2}&+&\frac{\lambda}{16 L}
(\sum\frac{1}{\omega})(\sum\frac{1}{\tilde{\omega}})+(\frac{\lambda}{32}-\frac
{\lambda}{16})\frac{1}{L}(\sum\frac{1}{\tilde{\omega}})^{2}\nonumber\\
+\frac{\lambda}{8mL^{2}} (\sum\frac{1}{\omega})^{2}&-&\frac{\lambda}{4mL^{2}}
(\sum\frac{1}{\omega})(\sum\frac{1}{\tilde{\omega}})+\frac{\lambda}{8mL^{2}}(
\sum\frac{1}{\tilde{\omega}})^{2}\nonumber\\
-\frac{\lambda m}{24 L^{2}}(\sum\frac{1}{\omega})(\sum\frac{1}{\omega^{3}})&+
&\frac{\lambda m}{24 L^{2}} (\sum\frac{1}{\omega^{3}})(\sum\frac{1}{\tilde{
\omega}})\nonumber\\
+\frac{\lambda m^{3}}{40 L^{2}}(\sum\frac{1}{\omega^{3}})^{2}&&\;\nonumber
\end{eqnarray}
\begin{eqnarray}
&&=-\frac{\lambda}{32 L}(\sum\frac{1}{\omega}-\sum\frac{1}{\tilde{\omega}})^{2}
+\frac{\lambda}{8m L^{2}}(\sum\frac{1}{\omega}-\sum\frac{1}{\tilde{\omega}})^
{2}\nonumber\\
&&-\frac{\lambda m}{24 L^{2}}(\sum\frac{1}{\omega^{3}})(\sum\frac{1}{\omega}
-\sum\frac{1}{\tilde{\omega}})+\frac{\lambda m^{3}}{40 L^{2}} (\sum \frac{1}{
\omega^{3}})^{2}
\end{eqnarray}
Drawn lines represent propagators in the soliton sector and dotted
lines
propagators in the trivial vacuum.
In the intermediate expression the first column gives the contributions from
the $\eta^{4}$ term in $H_{\rm int,sol}^{(2)}$ while the second column gives the 
contribution from $\delta m^{2}\eta^{2}$ in $\Delta H_{\rm sol}$. The third column
contains the contributions from the vacuum sector (in the first row) and from 
the term $(\delta m^{2})^{2}$ in the topological sector  (in the second row).
We recall that $\omega$ and $\tilde\omega$ denote the frequencies in
the topological and trivial sectors respectively.
We claim that all differences cancel. Since we already proved this for 
(\ref{dif}), 
we only need to discuss the sum $\frac{1}{L}(\sum\frac{1}{\omega}-\sum\frac{
1}{\tilde{\omega}})^{2}$. {\em Each factor} $\sum\frac{1}{\omega}-\sum
\frac{1}{\tilde{\omega}}$ {\em is ambiguous but finite}, see (\ref{dif}). 
Hence the extra $\frac{1}{L}$ factor ensures that also this term vanishes. 

We conclude that all $\eta^{4}, \eta^{2}$ and $\eta^{0}$ terms contribute 
only one term
\begin{equation}\label{129}
\frac{\lambda m^{3}}{40L^{2}}(\sum\frac{1}{\omega^{3}})^{2}=\frac{\lambda}{40
\pi^{2}m}
\end{equation}
This is the contribution due to the one-vertex two-loop graph in which
one only retains in both
propagators the deviations from the trivial space propagators.
The individual other terms are ambiguous and divergent but their sum cancels.

\subsubsection*{The $\eta^{3}$ and $\d m^2\eta$ contributions with one
intermediate particle} We next evaluate the contributions to the
mass of the soliton which come from the $\eta^{3}$ term in $H_{\rm
int,sol}^{(1)}$, and the $\eta$ term in $\Delta H_{\rm sol}$. We first
take one-particle intermediate states in the sums over $|p\rangle$ in
(1). Since there are no terms in the trivial sector which are odd in
$\eta$, these contributions should sum up to a finite result, but we
are again interested in possible ambiguities. Using

\begin{eqnarray}
&&\langle n|\frac{\delta m^{2}m}{\sqrt{\lambda}}\int_{-\infty}^{+\infty}\frac{-4(
e^{mx}-e^{-mx})}{(e^{mx}+e^{-mx})^{2}}\eta dx |{\rm sol}\rangle\nonumber\\
&&=-\frac{\delta m^{2}\pi}{m\sqrt{2L\lambda}}\frac{\sqrt{\omega_{n}}}{
\cosh \frac{1}{2}\pi q_{n}} 
\end{eqnarray}
(see the appendix), and
\begin{equation}
\frac{1}{L}\sum_{n}\frac{1}{ \cosh^{2}( \frac{1}{2} \pi q_{n} )}=\frac{2m}{
\pi^{2}}
\end{equation}
we find straightforwardly
\begin{equation}
\sum_{n}|\langle n|\eta\; {\rm term}|{\rm sol}\rangle|^{2}/(-\omega_{n})=-\frac{(\delta m^{2})^
{2}}{\lambda m}
\end{equation}

Next we evaluate $\langle p|\eta^{3}{\rm term}|{\rm sol}\rangle$ with $\langle p|$ a one-particle
state. This yields a factor of 3 times an equal-time contraction at the 
point $x$. The $x$-integrals are given in the appendix and one finds
\begin{equation}
\langle{\rm p,1\;part |\eta^{3}term|sol}\rangle=\sqrt{\frac{\lambda}{2L}}\frac{\pi}{4m}
\sum_{p}\frac{\sqrt{\omega_{p}}}{\cosh \frac{1}{2}\pi q_{p}}(\frac{1}{L}
\sum_{n}\frac{1}{\omega_n}-\frac{\omega_{p}^{2}}{4\pi m^{2}})
\end{equation}
Then it is relatively easy to obtain
\begin{eqnarray}
&&\sum_{p}\frac{2}{-\omega_{p}}Re\langle {\rm sol}|\eta^{3}{\rm term|p,1\;part\rangle\langle p,1\;part|
\eta \,term|sol}\rangle\nonumber\\&&
=(\delta m^{2})[-\frac{1}{2mL}\sum\frac{1}{\omega}-\frac{1}{6\pi m}]
\end{eqnarray}
Finally we evaluate
\begin{eqnarray}
&&\sum\frac{1}{-\omega_{p}}|\langle{\rm p,1\;part |\eta^{3}term |sol}\rangle|^{2}
\nonumber\\&&
=-\frac{1}{\lambda m}(\frac{\lambda}{4L}\sum\frac{1}{\omega})^{2}+\frac{1}
{6\pi m}(\frac{\lambda}{4L}\sum\frac{1}{\omega})-\frac{\lambda}{120 \pi^{2} m}
\end{eqnarray}

The sum of the contributions from the $\delta m^{2}\eta$ and $\eta^{3}$ terms
with only one-particle intermediate states is then
\begin{eqnarray}\label{136}
&&
\epsfysize 9pt\raisebox{4pt}{\epsfbox{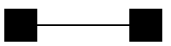}}
\qquad\raisebox{8pt}{$+$}\qquad
\epsfysize 16pt\epsfbox{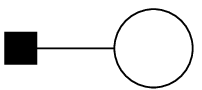}
\qquad\raisebox{8pt}{$+$}\qquad
\epsfysize 16pt\epsfbox{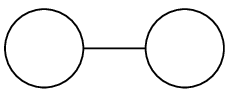}
\qquad\raisebox{8pt}{$=$}
\nonumber\\
&&
-\frac{1}{\lambda m}(\delta m^{2})^{2}+\delta m^{2}( \frac{1}{2 mL}\sum\frac{1}
{\omega}-\frac{1}{6\pi m})-\frac{1}{\lambda m}(\frac{\lambda}{4L}\sum\frac{1}{
\omega})^{2}\nonumber\\&&
+\frac{1}{6\pi m}(\frac{\lambda}{4L}\sum\frac{1}{\omega})-\frac{\lambda}{
120 \pi^{2} m}\nonumber\\&&
=-\frac{\l}{16 mL^{2}}(\sum\frac{1}{\omega}-\sum\frac{1}{\tilde{\omega}})^{2}+
\frac{\lambda}{24\pi mL}(\sum\frac{1}{\omega}-\sum\frac{1}{\tilde{\omega}})
\nonumber\\&&
-\frac{\lambda}{120\pi^{2}m}
\end{eqnarray}

The only nonvanishing contribution from these terms is thus due to the square
of the finite part of the matrix element of the $\eta^{3}$ term, and the latter
is again obtained by taking the deviation from the propagator in the trivial
vacuum (the part proportional to $\tanh^2 mx - 1$) in the equal-time contraction of two $\eta$ fields. All other matrix
elements are divergent, and none of them contributes.

\subsubsection*{The contribution from three intermediate particles and the
$\eta\pi\eta\pi$ term} From (\ref{H1I}) 
the matrix element with 3 intermediate particles can be
written as follows (after the substitution $e^{mx}=y$)
\eqa
&&\frac1{2m} \int_0^\infty dy^2 \biggl[ {(y^2-1)^4\over(y^2+1)^5} -i
{(y^2-1)^3\over(y^2+1)^4}(q_n+q_r+q_t)\nonumber\\
\qquad&&-{(y^2-1)^2\over(y^2+1)^3}(q_nq_r+q_nq_t+q_rq_t)+
i{y^2-1\over(y^2+1)^2}q_nq_rq_t \biggr](y^2)^{\frac{i}2Q-\frac12}
\eqae
where $Q=q_n+q_r+q_t$. Using
\eq
\int_0^\infty dy^2 y^n y^{iQ-1} (y^2+1)^{-m}=B(\frac{iQ}2+\frac{n+1}2,
\frac{-iQ}2-\frac{n+1}2+m)
\eqe
where $B$ is the beta function, one finds
\eqa
&&\frac1{2m}\left[ -\frac18 Q^4+\frac14 Q^2 + \frac12 (Q^2-1)
(q_nq_r+q_nq_t+q_rq_t)-Q(q_nq_rq_t)+\frac38 \right] {\pi\over\cosh\frac12\pi Q}
\nonumber\\
&=&\frac1{2m}\left[ -\frac18 (\omega_n^2+\omega_r^2+\omega_t^2)^2+
\frac12(\omega_n^2\omega_r^2+\omega_n^2\omega_t^2+\omega_r^2\omega_t^2)
\right] {\pi\over\cosh\frac12\pi Q}
\eqae
Squaring, one finds for the contributions to $M^{(2)}$ due to
intermediate states with 3 particles
\eqa
&&\frac16\sum_{n,r,t}\left| \langle q_n,q_r,q_t | H_{\rm
int,sol}^{(1),I} | {\rm sol} \rangle
\right|^2{-1\over\omega_n+\omega_r+\omega_t}
\qquad\raisebox{8pt}{$=$}\quad \epsfysize 20pt\epsfbox{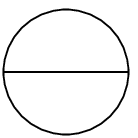}
\nonumber\\ &=&
{-\l\over2^{10}\pi 6m}\int{dq_1 dq_2 dq_3\over\cosh^2\frac12\pi(q_1+q_2+q_3)}
\nonumber\\&&\times
\textstyle{[(1+q_1^2)^2+(1+q_2^2)^2+(1+q_3^2)^2
-2(1+q_1^2)(1+q_2^2)-2(1+q_1^2)(1+q_3^2)-2(1+q_2^2)(1+q_3^2)]^2\over
(1+q_1^2)^{3/2}(1+q_2^2)^{3/2}(1+q_3^2)^{3/2}
(\sqrt{1+q_1^2}+\sqrt{1+q_2^2}+\sqrt{1+q_3^2})}
\eqae
The prefactor $\frac16$ is needed since we sum over all $q_n$, $q_r$, $q_t$
while the 3-particle states are given by $a^\dagger_n a^\dagger_r a^\dagger_t
|{\rm sol}\rangle$.

Next we evaluate the contribution from the $\int \eta'\pi\int \eta'\pi$ term.
Using that $\eta=\sum_n(2\omega_n L)^{-1}$ $(a_n E_n \exp -i\omega_n t + h.c.)$
while 
$\pi=\sum_n(2\omega_n L)^{-1}(-i\omega_na_n E_n \exp -i\omega_n t + h.c.)$
one finds 3 contributions from the 3 possible contractions,\footnote{
Each of these 3 contributions is proportional to $\hbar^2$ because
each propagator gives a factor $\hbar$, vertices do not contribute
factors of $\hbar$, and the energy denominators in (\ref{71}) yield a
factor $\hbar^{-1}$.}

\eqa
\langle {\rm sol}|H^{(2),III}_{\rm int,sol}|{\rm sol}\rangle=
\qquad\epsfysize 20pt\epsfbox{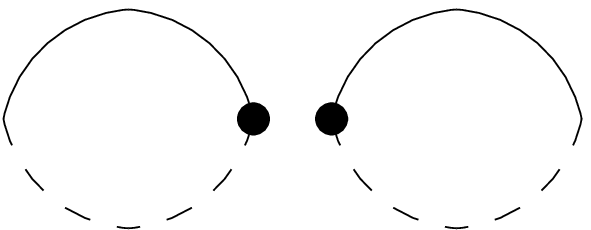}
\quad\raisebox{8pt}{$+$}\quad
\epsfysize 20pt\epsfbox{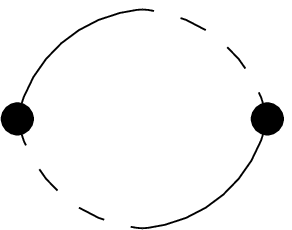}
\quad\raisebox{8pt}{$+$}\quad
\epsfysize 20pt\epsfbox{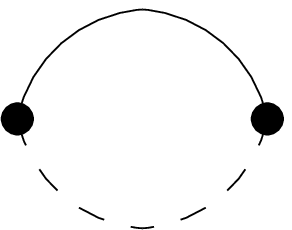}
\nonumber\\
=\frac\l{64m^3} \sum_{n,r}\left(-C_{n,n}C_{r,r}
+C_{r,-n}C_{-n,r}+{\omega_r\over\omega_n}C_{n,-r}C_{-n,r}\right)
\eqae
where $C_{n,r}=\int (dE_n/dx)E_r dx$. Dashed lines denote $\pi$
fields. Using that $E_n$ and $E_r$ are
orthonormal, it follows that
\eq
C_{n,r}=imq_n\d_{n,r}+\frac1L\frac1{\omega_n\omega_r}\left(q_n^2-q_r^2\right)
{i\pi\over\sinh\frac12\pi(q_n-q_r)}
\eqe
Since $C_{n,n}$ is odd in n $(C_{n,n}=imq_n+2iq_n/(L\omega_n^2))$
the contribution from $C_{n,n}C_{r,r}$ vanishes, and in the remainder
the term with $imq_n\d_{n,r}$ cancels in the contribution
$\sum_{r,n}({\omega_r\over\omega_n}-1)|C_{n,r}|^2$. This leads to
\eqa
&&\langle {\rm sol}|\int \eta'\pi\int \eta'\pi
\mbox{-term}|{\rm sol}\rangle\nonumber\\
&=&\frac\l{2^{10}m}\int{dq_1 dq_2\over[\sinh\frac12\pi(q_1-q_2)]^2}
\left( {\sqrt{1+q_1^2}\over\sqrt{1+q_2^2}}-1\right)
{(q_1^2-q_2^2)^2\over(1+q_1^2)(1+q_2^2)}
\eqae
According to Verwaest \cite{verw} the sum of these two contributions
is 
\begin{equation}\label{144}
-\frac\l{60\pi^2m}.
\end{equation}
(In ref. \cite{vega} these contributions were numerically evaluated).
For us the crucial point is that both contributions are finite and
hence unambiguous.

Adding (\ref{129}), (\ref{136}) and (\ref{144}) one finds that the sum
of these contributions vanishes. Hence, the only contribution to the
two-loop correction of the soliton mass comes from the term
$-\hbar^2/(8M_{\rm cl}^2)\int(\phi'')^2dx$ in (\ref{102}) and it
yields
\begin{equation}
M^{(2)} = -{\lambda\over192m}.
\end{equation}
All other contributions combine into unambiguous finite integrals
which vanish due to factors $1/L$.

\section{Conclusions}
\label{sec:conclude}

In this article two new concepts have been introduced in the theory of
quantum solitons: topological boundary conditions and a physical
principle which fixes UV quantum ambiguities.  The main idea
underlying these concepts is simple: the problem
must be formulated {\em before} the loop expansion
is performed. This means that the mass of the soliton
must be defined nonperturbatively, rather than as a sum of
the classical result, plus one-loop corrections, plus etc.
We define this mass as the difference between the vacuum
energies of the system with different boundary conditions.
It follows immediately that the boundary conditions must be formulated
for the full quantum field, rather than for small fluctuations.

The topological boundary conditions are, in fact, better viewed as
conditions which put the system on a Moebius strip without
boundaries. In the literature one usually employs periodic boundary
conditions in the soliton sector for all the quantum fluctuations (but
not for the classical field), or a mixture of periodic conditions for
bosons and others for fermions. All these conditions may distort the
system at the boundaries and introduce spurious extra energy $O(L^0)$
contributions which obscure the measurement of the mass of a
soliton. We even found that most boundary conditions in the literature
are not compatible with the Majorana condition for fermions in the
$N=(1,1)$ case. With topological boundary conditions there is no
spurious energy introduced at the boundaries since there are no
boundaries, and one obtains the genuine mass of a soliton.

In the long-standing problem of UV ambiguities of the quantum mass of
a soliton, we have taken the point of view that this problem should be
recognized as the well-known regularization dependence of loop
calculations in quantum field theory. The action is determined only up
to local counterterms, and one has to introduce renormalization
conditions to fix those. With solitons, we want to compare vacuum
energies in topologically distinct sectors.  The new principle we have
introduced is that if all mass parameters in the theory tend to zero,
the topological and the trivial sector must have the same vacuum
energy (in the infinite volume limit).

This condition follows immediately from a simple dimensional analysis.
Indeed, if the physical meson mass $m$ and all dimensionful
couplings scale according to their dimensions as $m\to0$, then
the mass of the soliton $M$ as a function of these parameters
must also scale as $m$. One can take a different, but
equivalent, look at the problem: After rotation to Euclidean
space one finds that the soliton is a string-like interface between
two phases in two-dimensional classical statistical field theory.
Our principle is then related to the well-known property
of the interface tension: the tension vanishes at the critical (conformal)
point~\cite{jdl}. Note again that the formulation of this principle
is entirely nonperturbative.

It is when we do the loop expansion we see that the principle leads to
nontrivial consequences. This is because divergent contributions
proportional to the UV cutoff $\Lambda$ and independent of $m$ may
appear (and they do appear in the $N=1$ theory with mode number
cutoff and topological boundary conditions). 
Our principle can be used as a renormalization condition to
eliminate such divergences.
Adopting this principle, the calculation of the soliton mass becomes
unambiguous if one first differentiates the sums over frequencies
w.r.t.\ the mass parameter, and then integrates setting the integration
constant equal to zero. 

There exist other
methods to compute masses in certain 1+1-dimensional models~
\cite{dhn2,poly,mccoy,are,kor,fad,scho,ahn}.
For exactly solvable models one can determine the S-matrix exactly
by assuming that the Yang-Baxter equation holds and that the S-matrix
factorizes into products of two-particle S-matrices.
This program has been extended to the $N=1$ susy sine-Gordon model
\cite{scho,ahn}
and from the result one reads off that to one loop
the soliton mass is given by
\begin{equation}
M=M_{\rm cl}(m) - {m\over2\pi},
\end{equation}
where $m$ is the physical renormalized meson mass.
This is the same result as we
have found with topological boundary conditions and our renormalization
condition at the conformal point. From our point of view, choosing the
Yang-Baxter equation (or the thermodynamic Bethe ansatz together with
the quantum Lax-pairs approach --- the ``inverse scattering method'')
amounts to a choice of regularization scheme, which, at least as far as
the quantum mass of solitons is concerned, appears to be equivalent
to our principle.

However, it may seem that the BPS bound $\langle
H\rangle\geq|\langle Z\rangle|$ is violated at the one-loop level
because of the negative sign of $M^{(1)}$, since the central charge
does not receive quantum corrections at one loop~\cite{imb1,rvn}
(apart from those absorbed by the renormalization of $m$).  We point
out that it is the unrenormalized expectation value $\langle
H\rangle$ of the Hamiltonian that should obey the bound,
not the physical soliton mass $M$, which may (and does in the $N=1$
case) differ from $\langle H\rangle$ by an $m$-independent counterterm.
Using the mode-number cut-off regularization and topological boundary
conditions, we have found that
\begin{equation}\label{HMNC}
\langle H \rangle_{\rm MNC} = M_{\rm cl}(m) + \frac\Lambda4 - {m\over2\pi}
\end{equation} 
where $\Lambda$ is the ultraviolet cut-off. This means that the bound
is observed by $\langle H\rangle$. This happens because $N=1$ susy is
not enough to eliminate the linear divergent term in $\langle
H\rangle$.  This term is positive in accordance with the BPS
bound. The bound is not saturated, which agrees with the observation 
of Olive and Witten \cite{WO} that the saturation of the bound is
related to ``multiplet shortening''. The latter does not occur in the
$N=1$ model.

Therefore, in the $N=1$ theory with mode number cutoff we encounter
a situation where our physical principle $M|_{m=0}=0$ leads
to nontrivial consequences. The subtraction of the trivial vacuum
energy as in (\ref{e1-e0}) eliminates bulk volume $O(L)$
contributions (in susy theories they are absent anyway). This subtraction
cannot, however, eliminate a possible $O(L^0)$ $m$-independent but
regularization dependent $O(\Lambda)$ contribution. The required
subtraction constant, or counterterm,
 is determined by the condition $M|_{m=0}=0$.
Therefore
\begin{equation}
M = \left( E_1 - E_0 \right) 
- \left( E_{1} - E_{0} \right)_{m=0}
\end{equation}
is a more complete (compared to (\ref{e1-e0}))
definition of the physical soliton mass. It is clear that this is
exactly the definition implemented by our $d/dm$ calculation
\begin{equation}
M = \int_0^m {d\over dm} ( E_{1} - E_{0} ).
\end{equation}

In $N=2$ models we do not encounter any of these issues. There it turned
out that neither the soliton mass nor the central charge receive quantum
corrections, hence the BPS bound remains intact and saturated.
In contrast with the
$D=2$ $N=1$ case where all susy representations,
with and without saturation of the bound, are
two-dimensional, the BPS bound of the $N=2$ models is protected
by ``multiplet shortening'' \cite{WO}.

An alternative UV regularization which one may use is the
energy cutoff. In the mode cutoff regularization one truncates
the {\em sums} over the modes. The energy cutoff amounts
to first converting the sums into integrals over momenta
and then truncating these {\em integrals}. As was shown in \cite{rvn}
this regularization scheme in the sine-Gordon model
leads to a result in disagreement with the Dashen-Hasslacher-Neveu
spectrum \cite{fad,dhn1}. In the supersymmetric sine-Gordon case
the energy cutoff would lead to a vanishing one-loop correction (after
standard renormalization of $m$), which is in contradiction with 
existing exact results~\cite{scho,ahn}. We examined
the two-loop corrections and found that no dependence on
the choice of regularization appears there. Therefore the difference
between the energy cutoff and the mode cutoff is purely a one-loop
effect. This suggests that, perhaps, a formulation of the theory exists
where this effect can be described in terms of a quantum topological
one-loop anomaly. Moreover, the one-loop
correction to the mass $M$ does not depend on the coupling constant,
thus it is, in a certain sense, a geometrical effect. 

\subsection*{Acknowledgments}
We would like to thank N. Graham and R. Jaffe for discussions.

\appendix

\section*{Appendix}

\begin{eqnarray}
\int_{-\infty}^{+\infty}[\frac{8}{(e^{mx}+e^{-mx})^{2}}]dx&=&\frac{4}{m}\\
\int_{-\infty}^{+\infty}[1-\frac{8}{(e^{mx}+e^{-mx})^{2}}](\tanh ^{2}mx -1
) dx&=&\frac{2}{3m}\\
\int_{-\infty}^{+\infty}[1-\frac{8}{(e^{mx}+e^{-mx})^{2}}](\tanh^{2}mx -1)^{2}
dx&=&-\frac{4}{5m}\\
\int_{-\infty}^{+\infty}\frac{e^{mx}-e^{-mx}}{(e^{mx}+e^{-mx})^{2}}
e^{i m q_{n}x}(\tanh mx-iq_{n}) dx
&=&\frac{\pi}{4m}\frac{1+q_{n}^{2}}{\cosh \frac{1}{2}\pi q_{n}}
\end{eqnarray}
\begin{equation}
\int_{-\infty}^{+\infty}\frac{e^{mx}-e^{-mx}}{(e^{mx}+e^{-mx})^{4}}e^{imq_{n}x}
(\tanh mx -iq_{n}) dx
=\frac{1}{64 m}(q_{n}^{2}+1)^{2}\frac{\pi}{\cosh \frac{1}{2}\pi q_{n}}
\end{equation}
(Substitute $e^{mx}=y$, write $iq_{n}e^{imq_{n}x}$ as $\frac{1}{m}\frac{d}{dx}
e^{imq_{n}x}$ and put $y^{2}=\frac{1}{t}-1$. Then use beta-functions and the identities $\Gamma (1+x)=x\Gamma (x)$ and $\Gamma (x)\Gamma (1-x)=\pi/\sin \pi x
$).

\begin{eqnarray}
\frac{1}{L}\sum_{p}\frac{1}{(\cosh \frac{1}{2}\pi q_{p})^{2}}&=&\frac{2m}
{\pi^{2}}\\
\frac{1}{L}\sum_{p} \frac{\omega_{p}^{2}}{(\cosh \frac{1}{2}\pi q_{p})^{2}}
&=&\frac{8m^{3}}{3\pi^{2}} \quad({\rm Use} \int_{0}^{\infty}\frac{xdx}{e^{\pi x}
+1}=\frac{1}{12}).\\
\frac{1}{L}\sum_{p}\frac{\omega_{p}^{4}}{(\cosh \frac{1}{2}\pi q_{p})^{2}}
&=&\frac{64 m^{5}}{15\pi^{2}}
\end{eqnarray}

\end{document}